\documentclass[smallextended,numbook]{myownsvjour3}

\usepackage{graphicx}
\usepackage{mathrsfs}
\usepackage{amssymb}
\usepackage{amsbsy}

\newcommand{\deriv}[2]{\frac{\mathrm{d}#1}{\mathrm{d}#2}}
\newcommand{\derivp}[2]{\frac{\partial #1}{\partial #2}}

\def\eqlaw{\stackrel{\mbox{\tiny (law)}}{=}} 
\renewcommand{\leq}{\leqslant}
\renewcommand{\geq}{\geqslant}
\newcommand{\AvParam}{\boldsymbol{\mu}}
\renewcommand{\a}{\alpha}
\newcommand{\abs}[1]{\left\vert#1\right\vert}
\newcommand{\ave}[1]{\overline{#1}}
\renewcommand{\b}{\beta}
\newcommand{\bm}{\boldsymbol{m}}
\newcommand{\comport}[2]{\,\mathrel{\mathop{#1}\limits_{#2}^{\null}}\,}
\renewcommand{\d}{{\mathrm d}}
\newcommand{\e}{{\mathrm e}}
\newcommand{\eps}{\varepsilon}
\newcommand{\g}{\gamma}
\newcommand{\half}{\frac{1}{2}}
\newcommand{\ii}{{\mathrm i}}
\newcommand{\ka}{\kappa}
\newcommand{\kk}{{\hat k}}
\newcommand{\la}{\lambda}
\newcommand{\mean}{{\mathbb E}}
\newcommand{\norm}[1]{\vert#1\vert}
\newcommand{\text}[1]{{\mathrm{#1}}}
\newcommand{\tr}{\mathop{\mathrm{tr}}\nolimits}
\newcommand{\z}{\zeta}
\newcommand{\Ai}{{\mathrm{Ai}}}
\newcommand{\Bi}{{\mathrm{Bi}}}
\newcommand{\C}{{\mathbf C}}
\newcommand{\D}{\Delta}
\newcommand{\E}{{\mathbf E}}
\newcommand{\F}{{{}_2{\mathrm F}_{1}}}
\newcommand{\G}{{\mathbf G}}
\newcommand{\GG}{{\hat G}}

\newcommand{\K}{{\mathbf K}}

\newcommand{\abar}{{\ave{\a}}}
\newcommand{\wbar}{{\ave{w}}}
\newcommand{\ubar}{{\ave{u}}}
\newcommand{\da}{{\delta\a}}
\newcommand{\dw}{{\delta w}}
\newcommand{\du}{{\delta u}}
\newcommand{\daa}{D_{\a\a}}
\newcommand{\daw}{D_{\a w}}
\newcommand{\dau}{D_{\a u}}
\newcommand{\dww}{D_{ww}}
\newcommand{\dwu}{D_{wu}}
\newcommand{\duu}{D_{uu}}
\newcommand{\hK}{{\widehat K}}
\newcommand{\hPhi}{{\widehat\Phi}}

\journalname{Journal of Statistical Physics}

\begin{document}

\title{The Lyapunov exponent of products of random $2\times2$ matrices close
to the identity}

\author{Alain Comtet \and Jean-Marc Luck \and Christophe Texier \and Yves
Tourigny}

\institute{
A. Comtet
\at
Univ. Paris Sud; LPTMS, UMR 8626, CNRS; 91405 Orsay cedex, France \\
Universit\'e Pierre et Marie Curie--Paris 6, 75005 Paris, France\\
\email{alain.comtet@u-psud.fr}
\and
J.M. Luck
\at Institut de Physique Th\'eorique, CEA Saclay and URA 2306, CNRS, 91191
Gif-sur-Yvette cedex, France\\
\email{jean-marc.luck@cea.fr}
\and
C. Texier
\at
Univ. Paris Sud; LPTMS, UMR 8626 and LPS, UMR 8502, CNRS; 91405 Orsay cedex,
France\\
\email{christophe.texier@u-psud.fr}
\and
Y. Tourigny
\at School of Mathematics, University of Bristol, Bristol BS8 1TW, United
Kingdom\\
\email{y.tourigny@bristol.ac.uk}
}

\date{\today}

\maketitle

\begin{abstract}
We study products of arbitrary random real $2 \times 2$ matrices that are close
to the identity matrix.
Using the Iwasawa decomposition of $\text{SL}(2,{\mathbb R})$,
we identify a continuum regime
where the mean values and the covariances of the three Iwasawa parameters
are simultaneously small. In this regime,
the Lyapunov exponent of the product is shown to assume a scaling form.
In the general case, the corresponding scaling function
is expressed in terms of Gauss' hypergeometric function.
A number of particular cases are also considered,
where the scaling function of the Lyapunov exponent
involves other special functions (Airy, Bessel, Whittaker, elliptic).
The general solution thus obtained allows us, among other things,
to recover in a unified framework many results
known previously from exactly solvable models of one-dimensional disordered
systems.
\keywords{Random matrices \and Iwasawa decomposition \and Lyapunov exponent
\and Disordered one-dimensional systems \and Quantum mechanics \and Anderson
localisation}
\end{abstract}

\section{Introduction}
\label{introductionSection}

Products of random matrices arise quite naturally
in the context of one-dimensional or quasi-one-dimensional
disordered systems~\cite{BL,cpv,Lu,pendry}.
Such products typically grow exponentially with the number of factors.
The associated growth rate $\g$
is called the {\em Lyapunov exponent} of the product of random matrices.

This paper deals with products of $2 \times 2$ random matrices with real
entries
and unit determinant, i.e., elements of $\text{SL}(2,{\mathbb R})$.
The relevance of products of $2\times 2$ transfer matrices
to one-dimensional disordered systems was already underlined
in the pioneering works by Dyson~\cite{Dy} and Schmidt~\cite{Sch}.
In the context of quantum mechanics in a one-dimensional random potential,
the exponential growth of random matrix products is at the basis
of the phenomenon of Anderson localization.
The Lyapunov exponent is interpreted as the inverse
of the localization length~\cite{Bor63}.
In classical statistical-mechanical models,
such as spin chains with random couplings and/or random fields,
the Lyapunov exponent is the reduced free energy per site~\cite{cpv,Lu}.
Products of random matrices have also received applications in physics
besides the one-dimensional realm.
To mention one example, in the context of turbulence
it has recently been pointed out that the Lyapunov exponent
characterizing the motion of inertial particles
transported by a turbulent flow can be computed by means of
a mapping onto a one-dimensional random potential~\cite{horvai,gawedzki}.

Besides perturbative weak-disorder expansions of various kinds,
numerical investigations,
and the Lloyd model, with its Cauchy distribution of random site energies,
which is solvable in any dimension~\cite{lloyd},
the cases where the Lyapunov exponent has been obtained in analytical
form are rather few, even in the simplest situation of $2\times 2$ matrices.
These exact solutions concern
classical disordered harmonic chains~\cite{Dy,theoh},
variants of the disordered Kronig-Penney model~\cite{theo,theopl},
the tight-binding Anderson model~\cite{BaLu},
and classical and quantum spin chains~\cite{NL,LN,FNT,LFN,jmspin}.
See also the monographs~\cite{BL,Lu,PF,CaLa} and,
for more recent examples,~\cite{MTW,CTT1,CTT2,CTT3}.
The success of the calculation in all those cases relies upon special properties
of the distribution of the random matrices,
with a single random variable whose distribution is simple
(i.e., exponential or power-law).

The goal of the present work is
to gain greater insight into the generic case by considering distributions
concentrated around the identity matrix. Our approach differs from perturbative
approaches such as those of Derrida {\em et al.}~\cite{dmp} and Sadel and
Schulz-Baldes~\cite{SSB}, in that we solve {\em exactly} a limiting form of the
integral equation for the Lyapunov exponent.
As we shall see, by using the Iwasawa decomposition of $\text{SL}(2,{\mathbb R})$
into compact, Abelian and nilpotent subgroups,
one can define a continuum regime in which the Lyapunov exponent exhibits a scaling form.
Our main finding is that the scaling form of the Lyapunov exponent can always
be expressed in terms of the logarithmic derivative of a special function
(Airy, Bessel, Whittaker, elliptic, hypergeometric).
The present approach can also be viewed as a systematic treatment
of the degenerate weak-disorder expansion put forward by Zanon and Derrida~\cite{zd}
(see Sec.~\ref{sec:zd}).
We thus recover in a unified framework many results
known previously from exactly solvable models of one-dimensional disordered systems,
and obtain several novel results.

\subsection{Contents of the paper}
\label{outlineSubsection}
The paper may be divided into four parts:
\begin{enumerate}
\item The first part is the remainder of this introduction, in which we give an
overview of our results and techniques. After some basic definitions and some
heuristics, we identify the continuum regime of interest,
and present the key equations; these are Equations~(\ref{ifp}) and
(\ref{transformedIfp})
for a certain invariant density and its Hilbert transform in the continuum
regime,
and Formula~(\ref{simplifiedLyapunov}) for the Lyapunov exponent. The
introduction ends with
a concrete calculation that illustrates our general approach in a simple
particular case.
\item The second part, Sec.~\ref{basicSection}, contains a careful derivation
of the key equations.
\item The third part is the core of the paper.
In Sec.~\ref{sec:weak},
we show how the characteristic exponent ---essentially a complexification of
the Lyapunov exponent--- can be systematically expanded
in powers of the covariances of the Iwasawa parameters. We also
compute the first two non-trivial terms of this weak-disorder expansion.
The following sections are concerned with the exact calculation
of the characteristic exponent in several situations, in order of increasing
difficulty.
Sec.~\ref{sec:mono} is devoted to the cases where only one
of the Iwasawa parameters is random;
for physical reasons that will emerge, the compact, Abelian and nilpotent cases
that arise are referred to as {\em distance}, {\em supersymmetric} and {\em
scalar} disorder
respectively.
The case
where there is both scalar and supersymmetric disorder, but no distance
disorder,
is treated in Sec.~\ref{sec:pot}.
We consider in Sec.~\ref{sec:ell} the situation where the three parameters
are uncorrelated and have zero mean.
Sec.~\ref{sec:gene} deals with the completely general case,
where the characteristic exponent may be expressed
in terms of Gauss's hypergeometric function.
The case of a vanishing stationary current
and the connection with the work of Zanon and Derrida~\cite{zd}
are also addressed there.
\item The fourth part brings out the relationships
between our results and various aspects of the theory of disordered systems.
Sec.~\ref{sec:HyperbolicBM} develops the connection with
Brownian motion in the
Poincar\'e hyperbolic half-plane.
Sec.~\ref{sec:MixedSchrodinger} discusses the interpretation of our results
in terms of a disordered quantum-mechanical model.
Finally, our findings are briefly summarised in Sec.~\ref{sec:disc},
while an appendix is devoted to the derivation of Equation~(\ref{transformedIfp}).
\end{enumerate}

\subsection{The Furstenberg formula}
\label{furstenbergSubsection}

We address the problem of computing the Lyapunov exponent
\begin{equation}
\gamma := \lim_{n \to \infty} \frac{{\mathbb E} \left ( \ln \left |
\Pi_n \right | \right )}{n}
\label{lyapunovExponent}
\end{equation}
of the product
\begin{equation}
\Pi_n := M_{n} M_{n-1} \cdots M_1,
\label{productOfMatrices}
\end{equation}
where the $M_n$ are independent random matrices in
$\text{SL}(2,{\mathbb R})$ with the same probability measure $m$. Here,
${\mathbb E} (\cdot)$ stands for the expectation with respect to~$m$,
and the notation $| \cdot |$ refers to the Euclidean norm in ${\mathbb R}^2$
and to the norm on $2 \times 2$ matrices induced by it.

Our starting point is the Furstenberg formula~\cite{BL,ChLe,Fu}:
\begin{equation}
\gamma = \int_{-\infty}^\infty p ( \d z ) \int_{\text{SL} \left
(2,{\mathbb R} \right )} m (\d M) \ln \frac{\left | M \pmatrix{z \cr 1}
\right |}{ \left | \pmatrix{ z \cr 1} \right |}.
\label{furstenbergFormula}
\end{equation}
In this expression,
the real variable $z$ appears as a {\it projective coordinate},
i.e., the reciprocal of the slope associated with a direction in ${\mathbb R}^2$,
and $p$ is the probability measure on the projective line (the set of all directions)
which is invariant under the action of matrices drawn from
$m$. We emphasize that, whereas $m$ may be considered as given, one essential
difficulty of the calculation is that $p$ must be {\em found}.
In the particular case where $p$ has a density, i.e.,
\begin{equation}
p (\d z) = f(z) \,\d z,
\end{equation}
it may be shown that the unknown density obeys the integral
Furstenberg equation ---often referred to as the ``Dyson-Schmidt equation'' in
the physical literature:
\begin{equation}
f(z) = \int_{\text{SL} \left ( 2, {\mathbb R} \right )} m (\d M ) \left ( f
\circ {\mathcal M}^{-1} \right ) (z)\,
\frac{\d {\mathcal M}^{-1}(z)}{\d z},
\label{furstenbergEquation}
\end{equation}
where ${\mathcal M}^{-1} (z)$ is the inverse of the Moebius
(linear fractional) transformation
\begin{equation}
{\mathcal M} (z) := \frac{m_{11} z + m_{12}}{m_{21} z +m_{22}}
\label{mtfdef}
\end{equation}
associated with the matrix
\begin{equation}
M := \pmatrix{ m_{11} & m_{12} \cr m_{21} & m_{22} }.
\label{mdef}
\end{equation}
Throughout this paper we consistently use the following notations:
\begin{itemize}
\item[$\bullet$]
Italic symbols ($M$, $T$) denote $2\times2$ matrices (see~(\ref{mdef}));
\item[$\bullet$]
Calligraphic symbols (${\mathcal M}$, ${\mathcal T}$) denote the corresponding Moebius
(linear fractional) transformations (see~(\ref{mtfdef}));
\item[$\bullet$]
Script symbols (${\mathscr M}$, ${\mathscr T}$) denote the corresponding linear operators
acting on functions (see~(\ref{eq:RepresentationOfSL2R})).
\end{itemize}

As pointed out earlier, there is no systematic method for solving
(\ref{furstenbergEquation});
exact solutions are limited to very specific forms of the distribution $m$
of the random matrix $M$.

\subsection{A one-dimensional disordered system}
\label{disorderedSubsection}

Our approach builds to a large extent on
the intimate connections between our topic and the theory
of one-dimensional disordered systems. In this subsection, we describe a simple
quantum-mechanical model that brings out the most relevant of
these connections.

Consider the Schr\"{o}dinger equation on the positive half-line
\begin{equation}
- \psi''(x) + V(x)\, \psi(x) = E\,\psi(x),
\label{schroedingerEquation}
\end{equation}
where the potential $V$ consists of arbitrary point scatterers
located at the points
\begin{equation}
0 < x_1 < x_2 < \cdots
\end{equation}
This means that the potential vanishes everywhere,
except at the positions~$x_n$ of the scatterers.
The action of each scatterer on the wave function $\psi$
is encoded in the linear rule
\begin{equation}
\pmatrix{
\psi' (x_n^+) \cr
\psi (x_n^+)
}
= B_n \pmatrix{
\psi' (x_n^-) \cr
\psi (x_n^-)
}
\label{scatterer}
\end{equation}
which relates the values of $\psi(x)$ and of its derivative $\psi'(x)$
to the left and to the right of the scatterer.
The requirement that the resulting Schr\"{o}dinger operator have a self-adjoint
extension
translates into the condition~\cite{ADK,Se}
\begin{equation}
\e^{\ii \theta} B_n \in \text{SL}(2,{\mathbb R})
\end{equation}
for some $\theta \in {\mathbb R}$, and there is no appreciable loss of
generality in taking $\theta=0$.
We then have
\begin{equation}
\label{eq:TransferMatrixForPsi}
\pmatrix{
\psi' (x_{n+1}^-) \cr
\psi (x_{n+1}^-)
}
= \Pi_n \pmatrix{
\psi' (x_1^-) \cr
\psi (x_1^-)
},
\end{equation}
where the matrices in the product are given by
\begin{equation}
\label{waveFunction}
M_n = \pmatrix{
\sqrt{k} & 0 \cr
0 & \frac{1}{\sqrt{k}}
}
\pmatrix{
\cos\a_n&-\sin\a_n\cr
\sin\a_n&\cos\a_n
}
B_n
\pmatrix{
\frac{1}{\sqrt{k}} & 0 \cr
0 & \sqrt{k}
}
\end{equation}
and belong to $\text{SL} \left ( 2, {\mathbb R} \right )$. Here, $k = \sqrt{E}$
is the momentum of the particle, $\a_n=k\ell_n$,
and $\ell_n = x_{n+1}-x_n$ is the distance between neighbouring scatterers.
The most familiar example is that of the standard (scalar) delta-scatterer
introduced
in the deterministic case by Kronig and Penney~\cite{KP}, and studied in the
random case by Frisch and Lloyd and others~\cite{LGP,FL}:
\begin{equation}
B_n = \pmatrix{
1 & u_n \cr
0 & 1
}.
\label{deltaScatterer}
\end{equation}

Frisch and Lloyd investigated the density of states of this model
by making use of the {\em Riccati variable}
\begin{equation}
z(x) := \frac{\psi'(x)}{\psi(x)},
\end{equation}
viewed as representing the ``position'' at ``time'' $x$ of a particle
moving along the real axis. They observed that the integrated density of states
per unit length is the (negative of the) stationary current of particles or,
what is the same,
the reciprocal of the
mean time that the Riccati variable takes to make a complete journey along
the real axis.

In his later study of the Frisch-Lloyd model, Halperin~\cite{Ha} made two
further contributions.
Firstly, he considered the particular limit of the model where the scatterers
become infinitely
dense whilst their strength becomes infinitely weak; we call this the {\em
continuum regime}. Secondly, Halperin recognised that it was not strictly
necessary to know the stationary distribution of the Riccati variable in order
to find the integrated density of states; he showed
how it could be obtained indirectly by considering
a certain {\em transform} of the stationary density.
While these early works were concerned with the integrated density of states,
it was eventually recognised that the Lyapunov exponent
and the integrated density of states are tightly related.
They can indeed be viewed as the real and imaginary parts
of a function analytic in the complex energy plane ---a property that
results in the Herbert-Jones-Thouless formula~\cite{HJ,Th}.

The approach and the results we present in this paper can be thought of as
extensions of Halperin's ideas to the case
where the delta-scatterer (\ref{deltaScatterer}) is replaced by the more
general point-scatterer
\begin{equation}
B_n = \pmatrix{
\e^{w_n} & 0 \cr
0 & \e^{-w_n}
}\,\pmatrix{
1 & u_n \cr
0 & 1
},
\label{doubleImpurity}
\end{equation}
with a scalar component of intensity $u_n$
and a supersymmetric component of intensity $w_n$.
Comtet {\em et al.} called this point-scatterer
the {\em double impurity}~\cite{CTT1}.
The relevance of our results to this quantum-mechanical model will be discussed
in Sec.~\ref{sec:MixedSchrodinger}.

\subsection{The complex characteristic exponent}
\label{characteristicSubsection}

Our first task is to identify the quantity that is, for the general product of
matrices $\Pi_n$, the counterpart of the integrated density of states (or
current) in the disordered quantum model of the previous subsection.
To this end, we integrate~(\ref{furstenbergEquation}) with respect to $z$:
\begin{equation}
\mean \left ( \int_z^{{\mathcal M}^{-1}(z)} f(t) \,\d t \right ) = j
\label{integralEquation}
\end{equation}
(using the notations set below~(\ref{mdef})).
The constant of integration $j$ is completely determined by
the requirement that $f$ be an invariant probability density.

We can gain some insight into the significance of the quantity $j$
by introducing a ``Riccati process'' $\{ z_n\}$ defined by the random
recurrence relation
\begin{equation}
\label{zrec}
z_{n+1}={\mathcal M}_{n+1}(z_n).
\end{equation}
Denote by $f_n$ the probability density of the random variable $z_n$. It then
follows easily from this recurrence relation that
\begin{equation}
\label{IntegralEqFn}
f_{n+1}(z)
= \mathbb{E} \left(
\frac{\d {\mathcal M}^{-1}(z)}{\d z}
\left [ f_n \circ {\mathcal M}^{-1} \right ] (z)
\right)
\end{equation}
and, after subtracting $f_n(z)$ from both sides,
\begin{equation}
\label{eq:FPlike}
f_{n+1}(z) -f_n(z) =
\mathbb{E} \left(
\deriv{\mathcal{M}^{-1} (z)}{z}\,
\left [ f_n \circ \mathcal{M}^{-1} \right ] (z)
- f_n(z)
\right).
\end{equation}
This discrete-time Fokker-Planck equation expresses the fact that
the iteration~(\ref{zrec}) leads to a redistribution of the probability density
of the projective coordinate $z$
(see below~(\ref{furstenbergFormula})).
By introducing the quantity
\begin{equation}
\label{eq:17}
j_n(z) := \mathbb{E}
\left(
\int_z^{\mathcal{M}^{-1}(z)} f_n(t)\,\d t
\right),
\end{equation}
we can write~(\ref{eq:FPlike}) in the ``conservation form''
\begin{equation}
\label{eq:Conservation}
f_{n+1}(z) -f_n(z) = \derivp{}{z}j_n(z).
\end{equation}
Therefore $j_n(z)$ has an obvious interpretation as (the negative of) the
``probability current'' induced by iteration.
Under mild assumptions on the distribution $m$ of the matrices in the product,
Furstenberg's theory asserts the existence of a stationary distribution
of the projective process and of a corresponding steady current:
\begin{equation}
f(z)=\lim_{n\to\infty}f_n(z) \quad \text{and} \quad j= \lim_{n\to\infty}j_n(z).
\end{equation}

This motivates the following choice for the definition of the
{\em complex characteristic exponent} associated with the product $\Pi_n$:
\begin{equation}
\fbox{$\displaystyle
\Omega := \gamma + \ii \pi j.
$}
\label{omegadef}
\end{equation}
We recall that the real and imaginary parts of $\Omega$ have
simple physical interpretations in the case of disordered quantum-mechanical problems:
the Lyapunov exponent $\gamma$ is the inverse of the localization length,
while the current $j$
identifies with the integrated density of states per scatterer,
sometimes also referred to as the rotation number.
Another physical interpretation of the Lyapunov exponent
can be found in the context of classical statistical-mechanical models
in one dimension,
where the Lyapunov exponent is the reduced free energy per site~\cite{cpv,Lu}.

\subsection{Decompositions of $\text{SL}(2,{\mathbb R})$ and the continuum
regime}
\label{continuumSubsection}

Our next task is to define a useful continuum limit for the product $\Pi_n$.
We remark that, for $k=1$, the disordered
quantum system introduced earlier expresses the matrix $M$ as the product
of a rotation matrix, a diagonal matrix and an upper triangular matrix:
\begin{equation}
M = \pmatrix{ \cos \a & -\sin \a \cr \sin \a & \cos \a }\,
\pmatrix{ \e^{w} & 0 \cr 0 & \e^{-w} } \,
\pmatrix{ 1 & u \cr 0 & 1 }.
\label{IwasawaDecomposition}
\end{equation}
By applying the Gram-Schmidt orthonormalisation algorithm to the column vectors
\begin{equation}
\bm_1=\pmatrix{m_{11}\cr m_{21}},\quad
\bm_2=\pmatrix{m_{12}\cr m_{22}}
\end{equation}
of the matrix $M$ parametrized as in~(\ref{mdef}),
it is easily shown that {\em every} element~$M$ of $\text{SL}(2,{\mathbb R})$
may be written in the form~(\ref{IwasawaDecomposition}).
Explicit formulae are as follows:
\begin{equation}
\pmatrix{\cos\a\cr\sin\a}=\frac{\bm_1}{\abs{\bm_1}},\quad
\e^w=\abs{\bm_1},\quad
u=\frac{\bm_1\cdot\bm_2}{\abs{\bm_1}^2}.
\label{GS}
\end{equation}

This is a particular instance of the {\em Iwasawa decomposition} of
a semi-simple Lie group into compact, Abelian and nilpotent
subgroups~\cite{iwasawa}.
Other familiar decompositions
\begin{equation}
M = T_1 (t_1) \,T_2 (t_2) \,T_3 (t_3)
\label{genericDecomposition}
\end{equation}
into one-parameter subgroups
are the Gauss decomposition, where $T_1$ is lower triangular, $T_2$ diagonal
and $T_3$ upper triangular,
and the Cartan decomposition, where $T_1$ and $T_3$ are rotation matrices, and
$T_2$ is diagonal.
Later in the paper, we shall work exclusively with the Iwasawa decomposition.
Nevertheless,
there is some merit in explaining in some generality how a decomposition such
as (\ref{genericDecomposition})
determines a certain continuum limit; see for instance~\cite{NS} for a study of
the ``Brownian bridge'' limit of a product of random matrices where the Cartan
decomposition
is put to good use.

Let us assume, then, that, for $i=1,2,3$, the matrix $T_i(t_i)$ is a
one-parameter subgroup such that $T_i(0)$ is the identity.
We choose as representation space the set of functions on the projective space
(see below~(\ref{furstenbergFormula})).
Then the operator
${\mathscr T}_i (t_i)$ acting on functions via
\begin{equation}
\label{eq:RepresentationOfSL2R}
{\mathscr T}_i (t_i)
f(z) := \left [ f \circ {\mathcal T}_i^{-1} (t_i) \right ] (z)
\end{equation}
is a representation of the subgroup
(using the notations set below~(\ref{mdef})).
The corresponding infinitesimal generator is
\begin{equation}
\label{eq:GeneratorD}
{\mathscr D}_{i} :=
\lim_{t_i \to 0} \frac{{\mathscr T}_i (t_i)- {\mathscr T}_i(0)}{t_i}
= g_i(z) \frac{\d}{\d z},
\end{equation}
where $g_i$ is a polynomial in $z$ of degree at most $2$ whose precise form
depends on the details of the subgroup.

In our context, the parameters $t_i$ are random variables;
we use the notation
\begin{equation}
\overline{t_i} := \mean(t_i),\quad
\delta t_i = t_i - \overline{t_i},\quad
\text{and} \quad D_{ij} = {\mathbb E} \left ( \delta t_i \,\delta t_j \right ),
\label{covariances}
\end{equation}
so that we have, in particular,
\begin{equation}
t_i = \overline{t_i} + \delta t_i.
\label{barNotation}
\end{equation}
It will be convenient to collect the means of the
parameters into a vector and the
covariances into a matrix:
\begin{equation}
\AvParam := \pmatrix{ \overline{t_1} \cr \overline{t_2} \cr \overline{t_3} }
\quad\text{and}\quad
{\boldsymbol \sigma}^2 := \pmatrix{
D_{11} & D_{12} & D_{13} \cr
D_{21} & D_{22} & D_{23} \cr
D_{31} & D_{32} & D_{33}
}.
\label{vecmat}
\end{equation}
The covariance matrix is symmetric and non-negative; its square root
${\boldsymbol \sigma}$ is therefore well-defined.

The continuum regime is defined as the scaling regime
where {\it all the expected values and the covariances are simultaneously
small}.
The integral equation (\ref{integralEquation})
for the unknown density $f$ may then be approximated by expanding the integral
on the left-hand side in powers of the random parameters,
neglecting terms of order greater than two, and
taking expectations.
The full derivation will be presented in
Sec.~\ref{detailedContinuumSubsection}.
This yields the following first-order differential equation for $f$:
\begin{equation}
\fbox{$\displaystyle
\frac{\d}{\d z} \left [ \frac{\sigma^2(z)}{2} \,f(z) \right ]-{v}(z)\,f(z)=j.
$}
\label{ifp}
\end{equation}
The local variance $\sigma^2(z)$ and the local velocity $v(z)$ appearing in
this equation are given respectively by
\begin{equation}
\sigma^2(z) := \left | \boldsymbol{\sigma}\,{\mathbf g}(z) \right |^2
\label{diffusionCoefficient}
\end{equation}
and
\begin{equation}
v(z) := \frac{1}{2} \,\left \{ {\mathbf g}'(z) \cdot \,\boldsymbol{\sigma}^2
{\mathbf g}(z) - {\mathbf c} \cdot \left [ {\mathbf g}(z) \times {\mathbf
g}'(z) \right ] \right \}
- \AvParam \cdot {\mathbf g}(z),
\label{driftCoefficient}
\end{equation}
where
${\mathbf c}$ is the ``correlation vector'' given by
\begin{equation}
{\mathbf c} := \pmatrix{
D_{23} \cr
-D_{13} \cr
D_{12}
}
\label{correlationVector}
\end{equation}
and ${\mathbf g}(z)$ is the vector-valued function with $i$th component
$g_i(z)$.
We shall refer to (\ref{ifp}) as the {\em integrated
Fokker-Planck equation} associated with the product of matrices.

\subsection{A stochastic differential equation}

The continuum regime introduced above also makes sense in the non-stationary
case.
The continuum limit of~(\ref{eq:FPlike}) takes the form of
the usual Fokker-Planck equation with $z$-dependent local velocity and local
variance coefficients.
An equivalent, more striking expression of this passage from the discrete to
the continuous is
the Stratonovich stochastic differential
equation
\begin{equation}
\label{eq:SDE}
\deriv{z(x)}{x} =- \AvParam \cdot {\mathbf g(z(x))} - \frac{1}{2} \,{\mathbf c}
\cdot
\left [ {\mathbf g}(z(x)) \times {\mathbf g}'(z(x)) \right ]
+ {\mathbf g}(z(x)) \cdot {\boldsymbol\eta}(x),
\end{equation}
where the continuous variable $x$ replaces the discrete variable $n$, and
the continuous process $z(x)$ replaces the discrete process $\{z_n\}$ defined
by the random recursion (\ref{zrec}). Here,
the $i$th component of the vector $\boldsymbol{\eta}$
is a white noise $\eta_i(x)$ associated with the component $T_i (t_i)$ in the
decomposition
(\ref{genericDecomposition}), the correlations between the components are given
by
\begin{equation}
\mean\left( \eta_i(x) \,\eta_j (x') \right) =
D_{ij} \, \delta(x-x') \qquad (1 \le i,\,j \le 3)
\end{equation}
and, as before, the
details of the decomposition (Iwasawa, Gauss, Cartan, etc.) are encoded in the
vector ${\mathbf g}(z)$.

\subsection{The Iwasawa decomposition and the Hilbert transform}
\label{hilbertSubsection}
Let us now restrict our attention to the particular case where the {\em Iwasawa
decomposition} is used.
For the ordering $t_1 = \alpha$, $t_2 = w$ and $t_3 = u$, we have
\begin{equation}
{\mathbf g} (z) = \pmatrix{1+z^2 \cr -2z \cr -1}.
\label{eq:DefVectorG}
\end{equation}
We shall write $D_{\alpha \alpha}$ instead of $D_{11}$, $D_{\alpha w}$ instead
of $D_{12}$, and so on.

A ``frontal attack'' on the problem of computing the Lyapunov exponent would
consist of finding the invariant density, and then evaluating the multiple
integral in the Furstenberg formula (\ref{furstenbergFormula}).
In the continuum regime, the right-hand side of this formula for
the Lyapunov exponent reduces to a single integral with respect to the
projective variable $z$:
\begin{eqnarray}
\gamma = -\overline{w} + \dau
&+& \left ( \overline{\a} + 2 \daw \right )\, \int_{-\infty}^\infty z f(z)\,\d
z
\nonumber\\
&+& \frac{1}{2} \daa \, \int_{-\infty}^\infty z \frac{\d}{\d z} \left [
(1+z^2) f(z) \right ] \,\d z,
\label{simplifiedLyapunov}
\end{eqnarray}
where the integrals on the right-hand side are Cauchy principal value
integrals.
Although this is a considerable simplification,
the task of evaluating the characteristic exponent in terms of familiar
functions is still
quite daunting in the general case where the matrices $M_n$ do not commute
among themselves.

In this work we shall follow a different approach,
in which a central part is played by
the {\em Hilbert transform} $F$ of the density
\begin{equation}
F (y) := \int_{-\infty}^{\infty} \frac{f(z)}{y - z} \,\d z,
\label{densityTransform}
\end{equation}
where $y$ is a complex variable in the lower half-plane.
Either the above transform
---whose relevance can be traced back to Dyson's seminal paper~\cite{Dy}---
or its primitive
\begin{equation}
{\bf F} (y) := \int_{-\infty}^{\infty} \ln(y-z)\,f(z)\,\d z
\end{equation}
have already been instrumental
in the derivation of exact solutions
of the Furstenberg equation (\ref{furstenbergEquation})
in many situations~\cite{Lu,theoh,theo,theopl,BaLu,NL,LN,FNT,LFN}.

We proceed to explain how the introduction of the Hilbert transform
facilitates our task in the continuum limit.
Dividing the integrated Fokker-Planck equation (\ref{ifp}) by $(y-z)$ and
integrating over $z$
yields the following equation for $F$:
\begin{equation}
\fbox{$\displaystyle
\frac{\d}{\d y} \left [ \frac{\sigma^2(y)}{2} \,F(y) \right
]-{v}(y)\,F(y)=\Omega + \frac{D_{\alpha \alpha}}{2} (1+y^2) + \overline{\alpha}
y - \overline{w}.
$}
\label{transformedIfp}
\end{equation}
A detailed derivation of this key equation will be given
in Appendix~\ref{transformAppendix},
starting from the material exposed in Sec.~\ref{lyapunovSubsection}.

Equation~(\ref{transformedIfp}) for the Hilbert transform
thus has the {\em same homogeneous part}
as the integrated Fokker-Planck equation~(\ref{ifp}) for the density.
This property holds because the coefficients $\sigma^2(y)$ and $v(y)$ are polynomials
(see~(\ref{ddef})).
The crucial point, however, is that the quantity of interest
---the complex characteristic exponent~$\Omega$---
now appears on the right-hand side of~(\ref{transformedIfp}).

\subsection{Explicit form of the key equation~(\ref{transformedIfp})}
\label{explicitSubsection}

To proceed, it will be useful to work with a fully explicit form
of~(\ref{transformedIfp}).
For the Iwasawa decomposition, the local velocity and variance coefficients are
polynomials in
the Riccati variable $z$ with respective degrees 3 and~4:
\begin{eqnarray}
v(z)&=&-\abar(z^2+1)+2\wbar z+\ubar
\nonumber\\
&+&\daa z(z^2+1)+2\dww z
\nonumber\\
&-&4\daw z^2-2\dau z+2\dwu,
\nonumber\\
\sigma^2(z)&=&\daa(z^2+1)^2+4\dww z^2+\duu
\nonumber\\
&-&4\daw z(z^2+1)-2\dau(z^2+1)+4\dwu z.
\label{ddef}
\end{eqnarray}
Hence the differential equation~(\ref{transformedIfp})
for the Hilbert transform $F(y)$ is equivalent to
\begin{equation}
Q(y)F'(y)+R(y)F(y)=S(y)+2\Omega,
\label{feq}
\end{equation}
where the coefficients
\begin{equation}
Q(y)=\sigma^2(y),\qquad
R(y)=2\left[\sigma(y)\sigma'(y)-v(y)\right],
\label{qrres}
\end{equation}
and $S(y)$
are polynomials in the complex variable $y$ of degree 4, 3 and 2 respectively.
They are given explicitly
by the formulae
\begin{eqnarray}
Q(y)&=&
\daa(y^2+1)^2+4\dww y^2+\duu
\nonumber\\
&-&4\daw y(y^2+1)-2\dau(y^2+1)+4\dwu y
\label{qdef}
\end{eqnarray}
and
\begin{equation}
R(y)=R_0(y)+R_2(y),\qquad
S(y)=S_0(y)+S_2(y),
\end{equation}
where the polynomials with subscript $0$ are linear in the mean Iwasawa
parameters,
and those with subscript $2$ are linear in their covariances:
\begin{eqnarray}
R_0(y)&=&2(\abar y^2-2\wbar y+\abar-\ubar),
\nonumber\\
S_0(y)&=&2(\abar y-\wbar),
\nonumber\\
R_2(y)&=&2(\daa y^3-2\daw y^2+(\daa+2\dww)y-2\daw),
\nonumber\\
S_2(y)&=&\daa(y^2+1).
\label{pols}
\end{eqnarray}

\subsection{Outline of the method.
Towards a classification of one-dimensional continuous disordered models}
\label{solutionSubsection}

The differential equation~(\ref{feq}) can be solved in two steps.
First, the integrating factor ---the solution of the corresponding homogeneous
equation--- is given by
\begin{equation}
H(y)=\exp\left(-\int\frac{R(y)}{Q(y)}\,\d y\right).
\label{heq}
\end{equation}
Second, the solution of the full, inhomogeneous equation
can be obtained by ``varying the constant'': by setting
\begin{equation}
F(y)=K(y)H(y),
\label{FKH}
\end{equation}
Equation~(\ref{feq}) is reduced to
\begin{equation}
K'(y)=\frac{S(y)+2\Omega}{Q(y)H(y)}.
\label{keq}
\end{equation}
We thus obtain
\begin{equation}
K(y)=\int\frac{S(y)+2\Omega}{Q(y)H(y)}\,\d y.
\label{ksol}
\end{equation}

The function $F(y)$ that we seek has two distinguishing properties: firstly, it
is analytic in the lower half-plane. Secondly,
since it is the transform of a probability density, it has the asymptotic
behaviour
\begin{equation}
F(y) \approx 1/y \quad \text{as}\ |y| \rightarrow \infty
\label{asymptoticBehaviour}
\end{equation}
along any ray contained in the lower half-plane.\footnote{In this work
the notation $A(x)\approx B(x)$ as $x\to x_0$
means $\lim_{x\to x_0}(A(x)/B(x))=1$.}
These two properties are sufficient to determine
both the characteristic exponent $\Omega$
and the interval of integration in~(\ref{ksol}), i.e., the full function
$F(y)$.

Let us illustrate this all-important point with the concrete example
of our monolithic scalar disorder (see Sec.~\ref{sec:sca}).
Consider a product
of matrices of the form (\ref{IwasawaDecomposition}) where only the $u_n$ are
random. This product corresponds
to a disordered quantum system with a potential made up of delta scatterers
whose strengths are random, and the
continuum regime is effectively the limit studied by Halperin~\cite{Ha}.
The only non-zero parameters are $\abar$, $\wbar$, $\ubar$ and $\duu$.
Hence
\begin{equation}
Q(y)=\duu
\end{equation}
is a constant. Equation~(\ref{heq}) yields
\begin{equation}
H(y)=\e^{P(y)/\duu},
\end{equation}
where $P(y)$ is the polynomial
\begin{equation}
P(y)=-\frac{2}{3}\abar y^3+2\wbar y^2+2(\ubar-\abar)y,
\end{equation}
whilst Equation~(\ref{keq}) reduces to
\begin{equation}
K'(y)= \frac{2}{\duu} ( \abar y-\wbar+\Omega)\,\e^{-P(y)/\duu}.
\label{halperinKsol}
\end{equation}

\begin{figure}[!ht]
\vspace{7cm}
\includegraphics{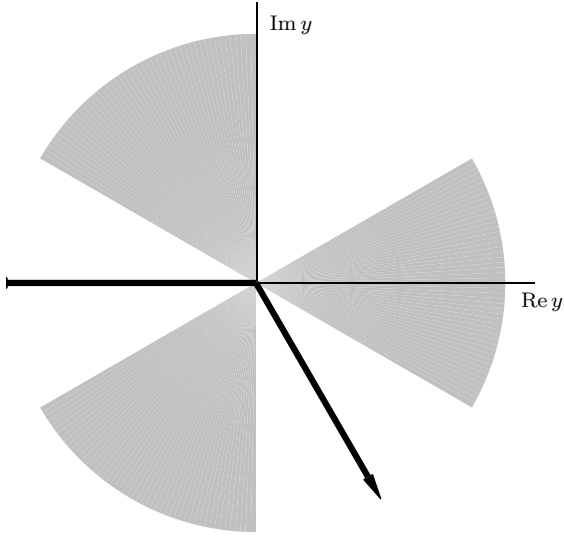}
\begin{picture}(0,0)
\put(60,94){\line(1,0){150}}
\put(205,85){$\text{Re} \,y$}
\put(105,95){\line(0,1){105}}
\put(110,190){$\text{Im} \,y$}
\end{picture}
\caption{Path of integration for monolithic distance disorder. The integrating
factor $H$ exhibits exponential growth at infinity in the unshaded sectors and
exponential decay in the
shaded ones.}
\label{airyFigure}
\end{figure}

Assume for definiteness that $\overline{\alpha}>0$. Then $H$ exhibits
exponential growth at infinity along any ray contained in the
unshaded sectors of the complex plane depicted in Figure~\ref{airyFigure}.
Since $F$ is analytic in the lower half-plane, Equation~(\ref{FKH}) and
the asymptotic condition
(\ref{asymptoticBehaviour}) together imply that $K$ must {\em decay to zero}
along any ray in the lower unshaded sectors.
Take for instance the piecewise linear path shown in Figure~\ref{airyFigure}:
it originates at $-\infty-\ii 0$ and follows the real axis in one of the
unshaded sectors until it reaches the origin,
where it is refracted and continues to infinity along the ray $\arg y = -\pi/3$
in an another unshaded sector. $K$ vanishes at the extremities of this
integration path; using (\ref{halperinKsol}) and the Fundamental Theorem of
Calculus, we deduce
\begin{equation}
\int_{-\infty}^{\e^{-\ii\pi/3}\infty}
(\abar y-\wbar+\Omega)\,\e^{-P(y)/\duu}\,\d y=0.
\label{cdai}
\end{equation}

The linear change of variable
\begin{equation}
y=\frac{\b t+\wbar}{\abar},
\end{equation}
where
\begin{equation}
\b=\left(\frac{\abar^2\duu}{2}\right)^{1/3} \quad \text{and}\quad
x=\frac{1}{\b^2} \left ( \wbar^2+\abar\,\ubar - \abar^2 \right
)=\frac{\mu^2}{\b^2}
\label{bxdef}
\end{equation}
(see~(\ref{musca})), brings the contour integral into the following form:
\begin{equation}
\int_{-\infty}^{\e^{-\ii\pi/3}\infty}
(\b t+\Omega)\,\e^{t^3/3-x t}\,\d t=0.
\end{equation}
The integrals involved can be expressed in terms
of the Airy function~\cite[(9.5.4), p.~196]{nist}
\begin{equation}
\Ai(x)=\int_{\e^{-\ii\pi/3}\infty}^{\e^{+\ii\pi/3}\infty}\frac{\d t}{2\pi\ii}\,
\e^{t^3/3-xt}
\end{equation}
and of its derivative with respect to $x$.
The end result is
\begin{equation}
\Omega=\b\,G(x),
\label{ressca}
\end{equation}
where the scaling function $G(x)$ is
\begin{equation}
\fbox{$\displaystyle
G(x)=\e^{-2\ii\pi/3}\frac{\Ai'(\e^{-2\ii\pi/3}x)}{\Ai(\e^{-2\ii\pi/3}x)}
=\frac{\Ai'(x)+\ii\,\Bi'(x)}{\Ai(x)+\ii\,\Bi(x)}.
$}
\label{gsca}
\end{equation}
This formula will be analysed in Sec.~\ref{sec:sca}.

In the general case, the complex zeros of the polynomial $Q(y)$
---which is nothing but the complexified diffusion coefficient---
will be at the heart of the subsequent analysis.
It is indeed clear from~(\ref{heq}) that the auxiliary function $H(y)$
is singular at the zeros of the polynomial $Q(y)$,
and that the multiplicities of the zeros dictate the nature of the
singularities.
For instance,
a simple zero at~$y_1$ yields the power-law singularity $H(y)\sim(y-y_1)^a$,
a double zero at $y_1$ yields the essential singularity $H(y)\sim(y-y_1)^a
\exp(b/(y-y_1))$,
and so on.\footnote{In this work
the notation $A(x)\sim B(x)$ as $x\to x_0$
is weaker than $A(x)\approx B(x)$.
It means that $A(x)$ and $B(x)$ become proportional as $x\to x_0$,
up to an unspecified factor,
which may either be constant or vary much more slowly than $A(x)$ or $B(x)$.}
It will be advantageous to take the view that the polynomial $Q(y)$ always has
four zeros,
counted with their multiplicities ---possibly including a zero at infinity.
Thus, for Halperin's example, we would say that $Q$ has a quadruple zero at infinity.
This convention is justified by the fact that the Riccati variable~$z$
is a projective coordinate (see below~(\ref{furstenbergFormula})).
Table~\ref{I} summarizes the relationships we have found between
types of disorder, patterns of zeros, and the special
functions that describe the scaling form of the characteristic exponent in the
continuum~regime.
As we are dealing with the most general matrix products of $\mathrm{SL}(2,\mathbb{R})$,
this table provides a classification
of one-dimensional continuous disordered models,
viewed as continuum limits of discrete models described by products of random matrices.

\begin{table}[!ht]
\begin{center}
\begin{tabular}{|c|c|c|c|c|}
\hline
Type of disorder & Ref. & Zeros & Special function & Sec. \\
\hline
\hline
Scalar & \cite{Ha} & 1q & Airy & \ref{sec:sca}\\
\hline
Supersymmetric & \cite{BCGL} & 2d & Bessel (real index) & \ref{sec:susy}\\
\hline
Distance & (new) & 2d & Bessel (imaginary index) & \ref{sec:dis}\\
\hline
General potential & \cite{HagTex08} & 1d+2s & Whittaker & \ref{sec:pot}\\
\hline
Independent (zero mean) & (new) & 4s & elliptic & \ref{sec:ell}\\
\hline
Most general & (new) & 4s & hypergeometric & \ref{sec:gene}\\
\hline
\end{tabular}
\end{center}
\caption{\small List of the examples worked out in this paper,
giving
the type of disorder
(with reference to earlier works, when applicable),
the pattern of zeros of the polynomial $Q(y)$
(q, d, s respectively denote quadruple, double and simple zeros),
the type of special function entering the scaling form
of the characteristic exponent $\Omega$,
and the section where the analysis is carried out.}
\label{I}
\end{table}

\section{Basic tools and derivation of the key formulae}
\label{basicSection}

In this section we present a detailed derivation of the key formulae
upon which our study relies, namely of Equations
(\ref{ifp}), (\ref{simplifiedLyapunov}) and (\ref{transformedIfp}).

\subsection{The equation for the density in the continuum regime}
\label{detailedContinuumSubsection}

We use the notations set below~(\ref{mdef}),
consistently with Sec.~\ref{continuumSubsection},
and set
\begin{equation}
\xi (t_1,t_2,t_3) := \int_z^{{\mathcal M}^{-1}(z)} \d t \, f(t)
= P \left( {\mathcal M}^{-1}(z) \right) - P (z),
\end{equation}
where $P$ is a primitive of $f$. Our aim is to determine the quadratic Taylor
polynomial of $\xi$ when the origin
is the point of expansion.
An elegant way of doing this is to introduce the representation ${\mathscr T}$
of $\text{SL} (2,{\mathbb R})$ which maps the
matrix
\begin{equation}
M = T_1 (t_1) \,T_2 (t_2) \,T_3(t_3)
\end{equation}
to the operator ${\mathscr T}_M$ defined by
\begin{equation}
{\mathscr T}_M := {\mathscr T}_1 (t_1) \circ {\mathscr T}_2 (t_2) \circ
{\mathscr T}_3 (t_3).
\end{equation}
Then
\begin{equation}
\xi (t_1,t_2,t_3) = \left ( {\mathscr T}_M P \right ) (z) - \left ( {\mathscr
T}_I P \right ) (z)
\label{xiRepresentation}
\end{equation}
where $I$ is the identity matrix.
The operator ${\mathscr T}_M$ may be expressed in terms of the subgroup
generators as
\begin{equation}
{\mathscr T}_M = \e^{t_1 {\mathscr D}_1} \,\e^{t_2 {\mathscr D}_2} \,\e^{t_3
{\mathscr D}_3}.
\end{equation}
Reporting this in~(\ref{xiRepresentation}) and expanding the exponential, we
obtain
\begin{eqnarray}
\label{eq:ExpansionXi}
\xi (t_1,t_2,t_3)=
\bigg[
t_1\, \mathscr{D}_1 + t_2\, \mathscr{D}_2 + t_3\,\mathscr{D}_3
+ \frac12 t_1^2\,\mathscr{D}_1 ^2
+ \frac12 t_2^2\,\mathscr{D}_2^2
+ \frac12 t_3^2 \mathscr{D}_3^2
\nonumber\\
+ t_1 t_2 \,\mathscr{D}_1\,\mathscr{D}_2
+ t_1 t_3 \,\mathscr{D}_1\,\mathscr{D}_3
+ t_2 t_3 \,\mathscr{D}_2\,\mathscr{D}_3 +\cdots
\bigg] P(z).
\end{eqnarray}
The identity $\mathscr{D}_i P(z)=f(z)\mathscr{D}_i z$
then leads to the following formulae for the partial derivatives of $\xi$:
\begin{equation}
\frac{\partial \xi}{\partial t_i} \Bigl |_{t_1=t_2=t_3=0} = f(z) {\mathscr D}_i
z
\end{equation}
and
\begin{equation}
\frac{\partial^2 \xi}{\partial t_i \partial t_j} \Bigl |_{t_1=t_2=t_3=0} =
\left\{
\begin{array}{ll}
{\mathscr D}_i \left [ f(z) {\mathscr D}_j z \right ] & \;\mbox{if}\ i \le j,
\\
& \\
{\mathscr D}_{i} \left [ f(z) {\mathscr D}_{j} z \right ] + f(z) \left [
{\mathscr D}_{i}, {\mathscr D}_{j} \right ] z\ \ & \;\mbox{if}\ i>j,
\end{array}
\right.
\label{hessian}
\end{equation}
where $[ {\mathscr D}_i, \,{\mathscr D}_j ]$ is the commutator of ${\mathscr
D}_i$ and ${\mathscr D}_j$. Replacing $\xi$
in~(\ref{integralEquation}) by its quadratic Taylor polynomial leads
to~(\ref{ifp}).

\subsection{Calculation of the invariant measure}
\label{invariantSubsection}

Although our approach does not require explicit knowledge of the invariant
density $f$, its
calculation is often of independent interest, and so we discuss it here. As a
by-product, we shall obtain information
on the large-$z$ behaviour of $f(z)$ that will prove useful in the derivation
of the equation for its Hilbert transform.

The solutions of the Fokker-Planck equation form a two-parameter family: the
current
$j$ is one of these parameters; the other is
a constant of integration.
We now explain how these two parameters
must be chosen in order
to yield the invariant measure. It will be helpful to
work with the {\em angle} variable
\begin{equation}
\varphi = 2 \arctan z \in [-\pi,\pi].
\end{equation}
Let us indicate very briefly the equations appropriate for this choice.
We shall use the subscript $a$ to distinguish the density $f_a$ of the angular
variable $\varphi$ from the
density $f$ of the projective variable $z$, and similarly for the angular
counterparts of other functions such as $\sigma^2$, $v$, ${\mathbf g}$, etc.
We can express the relationship between the densities as
\begin{equation}
(1+z^2) f(z) = 2\,f_{a} (\varphi).
\label{densities}
\end{equation}
In the continuum
regime, it is easy to see by substitution in (\ref{ifp})
that the density $f_a$ satisfies
\begin{equation}
\frac{\d}{\d \varphi} \left [\frac{ \sigma_a^2 (\varphi)}{2} f_{a}(\varphi)
\right ] - v_a(\varphi) f_{a}(\varphi) = j,
\label{integratedFokkerPlanck}
\end{equation}
where the local variance $\sigma_a^2$ and the local velocity $v_a$ are given by
formulae analogous to~(\ref{diffusionCoefficient})--(\ref{driftCoefficient}):
\begin{equation}
\sigma_a^2(\varphi) := \left | \boldsymbol{\sigma}\,{\mathbf g}_a(\varphi)
\right |^2
\label{angulaDiffusionCoefficient}
\end{equation}
and
\begin{equation}
v_a(\varphi) := \frac{1}{2} \,\left \{ {\mathbf g}_a'(\varphi) \cdot
\,\boldsymbol{\sigma}^2 {\mathbf g}_a(\varphi) - {\mathbf c} \cdot \left [
{\mathbf g}_a(\varphi) \times {\mathbf g}_a'(\varphi) \right ] \right \}
- \AvParam \cdot {\mathbf g}_a(\varphi),
\label{angularDriftCoefficient}
\end{equation}
where
\begin{equation}
{\mathbf g}_a (\varphi) = \pmatrix{
2 \cr
-2 \sin \varphi \cr
-1-\cos \varphi
}.
\label{angleGfunction}
\end{equation}
In the continuum regime, the angle variable is therefore a diffusion process
on the interval $[-\pi,\pi]$ with infinitesimal generator
\begin{equation}
{\mathscr G} := \frac{\sigma_a^2 (\varphi)}{2} \frac{\d^2}{\d \varphi^2} +
v_a(\varphi)
\frac{\d}{\d \varphi}
\label{infinitesimalGenerator}
\end{equation}
acting on the space of twice-differentiable $2 \pi$-periodic functions.

Set
\begin{equation}
J(\varphi) := 2 \int_{-\pi}^{\varphi} \frac{v_a(t)}{\sigma_a^2(t)} \,\d t.
\label{currentFunction}
\end{equation}
Equation~(\ref{integratedFokkerPlanck}) may be integrated to yield
\begin{equation}
f_a(\varphi) = \frac{2\,\e^{J(\varphi)}}{\sigma_a^2(\varphi)} \left [ C + j
\,\int_{-\pi}^{\varphi} \e^{-J(t)} \,\d t \right ].
\end{equation}
The two parameters $C$ and $j$ in this expression are determined by requiring
that $f_a$ be normalised and $2 \pi$-periodic. The periodicity
condition yields
\begin{equation}
\left [ 1-\e^{J(\pi)} \right ] C = j\,\e^{J(\pi)} \,\int_{-\pi}^{\pi}
\e^{-J(\varphi)} \,\d \varphi.
\label{periodicityCondition}
\end{equation}
There are two cases to consider.

\begin{itemize}

\item[$\bullet$]
The first case arises when
\begin{equation}
J(\pi)=2\int_{-\pi}^\pi \frac{v_a(t)}{\sigma_a^2(t)} \,\d t=0.
\label{jzero}
\end{equation}
Then the periodicity condition (\ref{periodicityCondition}) implies $j=0$.
The general situation where the stationary probability current vanishes
will be investigated in Sec.~\ref{vanishing},
while we shall encounter an interesting special case in
Sec.~\ref{sec:HyperbolicBM}.
For the time being, let us observe that the condition~(\ref{jzero})
is equivalent to
\begin{equation}
\int_{-\infty}^\infty \frac{v(z)}{\sigma^2 (z)}\,\d z=0.
\label{j0}
\end{equation}
The density is then given by
\begin{equation}
f _{a}(\varphi) = 2 C\, \frac{\e^{J(\varphi)}}{\sigma_a^2(\varphi)}
\label{zeroCurrentDensity}
\end{equation}
and $C$ is the normalisation constant.

\item[$\bullet$]
The second case corresponds to $J(\pi)
\ne 0$. Then the periodicity condition (\ref{periodicityCondition}) expresses
$C$ in terms of $j$, the density is given by
\begin{equation}
f_{a}(\varphi) =
\frac{2j}{1-\e^{J(\pi)}}\,\frac{\e^{J(\varphi)}}{\sigma_a^2(\varphi)}\,\left [
\e^{J(\pi)}\int_{\varphi}^{\pi} \e^{-J(t)} \,\d t + \int_{-\pi}^{\varphi}
\e^{-J(t)} \,\d t \right ]
\label{nonzeroCurrentDensity}
\end{equation}
and $j$ plays the r\^{o}le of a normalisation constant.

\end{itemize}

\subsection{The Rice formula}
\label{riceSubsection}

By setting $\varphi = \pm \pi$ in (\ref{integratedFokkerPlanck}), we readily
obtain
\begin{equation}
4 \daa \,f_{a}'(\pm \pi) + 2 \left ( \overline{\a} + 2
\daw \right ) f_{a}(\pm \pi) = j.
\label{periodicRiceFormula}
\end{equation}
When $\daa > 0$, the periodicity of $f_{a}$ therefore
implies the
periodicity of (all) its derivative(s).
By expressing this periodicity in terms of the density $f$ of the Riccati
variable $z$, we find
\begin{equation}
\lim_{z \to -\infty} z^2 f(z) = \lim_{z \to \infty} z^2 f(z)
\label{rice1}
\end{equation}
and
\begin{equation}
\lim_{z \to -\infty} z^2 \frac{\d}{\d z} \left [ (1+z^2)\,f(z) \right ]
=
\lim_{z \to \infty} z^2 \frac{\d}{\d z} \left [ (1+z^2)\,f(z) \right
].
\label{rice2}
\end{equation}
Furthermore, (\ref{periodicRiceFormula}) becomes
\begin{equation}
\daa \lim_{|z| \to \infty} z^2 \frac{\d}{\d z} \left [
(1+z^2)\,f(z) \right ]
+ \left ( \overline{\a} + 2 \daw \right ) \lim_{|z|
\to \infty} z^2 f(z) = j.
\label{riceFormula}
\end{equation}
In particular, the case where $\a$ is non-random (i.e., $\daa=0$) yields
\begin{equation}
\lim_{|z| \to \infty} z^2 f(z) = \frac{j}{\abar}.
\end{equation}
We thus recover a well-known relationship
between the integrated density of states
and the decay of the stationary density at infinity~\cite{FL}.
The above identity can be viewed as a special case of a formula due to Rice
for the density of zeros of a continuous process~\cite{rice}.
We shall henceforth refer to (\ref{riceFormula}) as the {\em Rice formula}.

\subsection{The Lyapunov exponent in the continuum regime}
\label{lyapunovSubsection}

Following~\cite[Sec.~3.4]{CTT1}, we write
\begin{equation}
B := \pmatrix{\e^w & 0 \cr
0 & \e^{-w}
} \, \pmatrix{ 1 & u \cr
0 & 1
}
\end{equation}
so that the Furstenberg formula (\ref{furstenbergFormula}) for the Lyapunov
exponent may be expressed in the form
(see the bottom of p. 441 in~\cite{CTT1})
\begin{equation}
\gamma = -\overline{w} + \frac{1}{2} \,{\mathbb E}
\left ( \int_{-\infty}^\infty \d z \,\ln (1+z^2) \frac{\d}{\d z}
\int_z^{{\mathcal B}^{-1}(z)} \d t f(t)
\right ).
\end{equation}
By using integration by parts in the integral over $z$, we obtain
\begin{equation}
\gamma = -\overline{w} - {\mathbb E} \left ( \int_{-\infty}^\infty \d z
\,\frac{z}{1+z^2} \int_z^{{\mathcal B}^{-1}(z)} \d t f(t) \right ).
\label{CTTformula}
\end{equation}
This formula is valid whether or not one considers the continuum regime; we
shall now show how it simplifies in that regime.

Set
\begin{equation}
\chi := \int_{-\infty}^\infty \d z \,\frac{z}{1+z^2} \int_z^{{\mathcal
B}^{-1}(z)}
\d t f(t).
\end{equation}
Just as, in Sec.~\ref{detailedContinuumSubsection}, we expanded the integral
$\xi$ in powers of $\a$, $w$ and $u$, we can expand the integral
$\chi$
in powers of $w$ and $u$. After taking expectations, we find
\begin{equation}
\gamma = -\overline{w} - \int_{-\infty}^\infty \d z \,\frac{z}{1+z^2} \,h \left
( \overline{w},\overline{u},\dww,\dwu,\duu;z \right ),
\label{intermediateLyapunov}
\end{equation}
where
\begin{eqnarray}
h &:=& \left ( \overline{w} {\mathscr D}_{w} z + \overline{u}
{\mathscr D}_u z
+ \frac{1}{2} \dwu \left [ {\mathscr D}_{w},\,{\mathscr
D}_{u}\right ] z \right )\, f(z)
\nonumber\\
&+&
\frac{1}{2} \Bigl \{ \dww {\mathscr D}_{w} \left [ f(z)
{\mathscr D}_{w} z \right ] + \dwu {\mathscr D}_{w}
\left [ f(z) {\mathscr D}_{u} z \right ]
\nonumber\\
&& {\hskip 7pt}+\dwu{\mathscr D}_{u} \left [ f(z) {\mathscr D}_{w}
z \right ] + \duu {\mathscr D}_{u} \left [ f(z) {\mathscr D}_{u} z
\right ] \Bigr \}.
\label{hFormula}
\end{eqnarray}

We then use the integrated Fokker-Planck equation
(\ref{ifp}) to express $h$ in the equivalent form
\begin{equation}
h = j - \overline{\a} \,f(z) {\mathscr D}_{\a} z - \frac{1}{2}
{\mathscr D}_{\a} \Bigl \{ f(z) \left [ \daa {\mathscr
D}_{\a} z
+ 2 \daw {\mathscr D}_{w} z + 2 \dau {\mathscr
D}_{u} z\right ] \Bigr \}.
\end{equation}
Here, the important observation is the occurrence of the operator ${\mathscr
D}_{\a}$ in all but the first term on the right-hand side. Since
\begin{equation}
{\mathscr D}_{\a} = (1+z^2) \frac{\d}{\d z},
\end{equation}
the awkward denominator in the integrand of (\ref{CTTformula}) will be removed.
Reporting this in (\ref{intermediateLyapunov}), we find
\begin{eqnarray}
\gamma &=&
-\overline{w}+\overline{\a}\, \int_{-\infty}^\infty z f(z)\,\d z \nonumber\\
&+&
\int_{-\infty}^\infty z\,\frac{\d}{\d z}
\left\{
f(z)\left[
\frac{\daa}{2}\, g_\a(z)
+ \daw\, g_w(z) + \dau\, g_u(z)
\right]
\right\} \d z,
\label{Lya}
\end{eqnarray}
where the $g_i(z)={\mathscr D}_{i} z$, $i\in\{\a,\,w,\,u\}$,
are the components of the vector (\ref{eq:DefVectorG}).

This formula may be simplified somewhat by performing an integration by parts
involving the last two terms in the square brackets;
this yields (\ref{simplifiedLyapunov}).
The derivation of (\ref{transformedIfp}) for the Hilbert
transform is now reasonably straightforward but somewhat
lengthy (see Appendix~\ref{transformAppendix}).

\section{The weak-disorder regime}
\label{sec:weak}

In this section we describe a systematic weak-disorder expansion of the characteristic
exponent in powers of the covariances.
Although the weak-disorder expansion is only a divergent asymptotic expansion in general
(see below~(\ref{nonpert})),
the knowledge of the first few terms has its own interest.
Besides this, it will also provide a useful means of checking the correctness
of the scaling forms that we shall determine later on in various special cases.

\subsection{No disorder}

We start by considering the case where there is no disorder.
In this situation,
all the matrices $M_n$ are equal to the constant matrix $M_0$
corresponding to $\abar$, $\wbar$ and $\ubar$.
The latter matrix is of the form~(\ref{mdef}) with
\begin{eqnarray}
&&m_{11}=\cos\abar\,\e^{\wbar},\qquad m_{12}
=\cos\abar\,\e^{\wbar}\ubar-\sin\abar\,\e^{-\wbar},
\nonumber\\
&&m_{21}=\sin\abar\,\e^{\wbar},\qquad m_{22}
=\cos\abar\,\e^{-\wbar}+\sin\abar\,\e^{\wbar}\ubar.
\end{eqnarray}
Let us introduce the variable $\mu$ such that
\begin{equation}
\tr M_0=\sin\abar\,\e^\wbar\ubar+2\cos\abar\cosh\wbar=2\cosh\mu.
\label{muave}
\end{equation}
The eigenvalues of $M_0$ are therefore $\e^{\pm\mu}$.
The recursion~(\ref{zrec}) reduces to the deterministic
Moebius transformation
\begin{equation}
z_{n+1}={\mathcal M}_0(z_n)
\label{map}
\end{equation}
whose fixed points are
\begin{equation}
z_\pm=\frac{\cos\abar-\e^{-(\wbar\pm\mu)}}{\sin\abar}.
\label{zpm}
\end{equation}
Let us assume for definiteness that $\mu$ has a non-zero real part,
and choose $\mu$ so that
\begin{equation}
\text{Re}\,\mu>0.
\label{mupos}
\end{equation}
If $\abs{\tr M_0}\le2$, so that~(\ref{muave}) yields imaginary values of $\mu$,
we supplement $\mu$ with an infinitesimal positive real part, so
that~(\ref{mupos})
still holds.
This condition ensures that $z_+$ (resp.~$z_-$) is the stable (resp.~unstable)
fixed point
of the mapping~${\mathcal M}_0$.
We thus obtain the simple result
\begin{equation}
f_0(z)=\delta(z-z_+)
\label{f0}
\end{equation}
for the invariant density of the Riccati variable,
and the expected expression
\begin{equation}
\Omega_0=\mu
\label{omega0}
\end{equation}
for the characteristic exponent.

In the continuum regime where $\abar$, $\wbar$ and $\ubar$ are small,
the above expressions~(\ref{muave}) and~(\ref{zpm}) respectively simplify to
\begin{equation}
\fbox{$\displaystyle
\mu=\sqrt{\wbar^2+\abar\,\ubar-\abar^2}
$}
\label{musca}
\end{equation}
and
\begin{equation}
z_\pm=\frac{\wbar\pm\mu}{\abar}=\frac{\abar-\ubar}{\wbar\mp\mu}.
\label{fpmap}
\end{equation}
Throughout the following,
$\mu$ will be a notation for the right-hand side of~(\ref{musca}).

Let us now examine the general formalism
in the present non-random situation.
The only non-zero polynomials are $R_0(y)$ and $S_0(y)$.
Equation~(\ref{feq}) therefore loses its differential character.
Using the subscript $0$ to indicate that there is no disorder, we
obtain
\begin{equation}
F_0(y)=\frac{S_0(y)+2\Omega_0}{R_0(y)}.
\end{equation}
The polynomial $R_0(y)$ factorizes as
\begin{equation}
R_0(y)=2\abar(y-z_+)(y-z_-),
\end{equation}
where $z_\pm$ are the fixed points (\ref{fpmap}) of the mapping~(\ref{map})
in the continuum limit.
The natural condition that $F_0(y)$ have no pole at the unstable zero ($y=z_-$)
allows one to recover the results~(\ref{f0}) and~(\ref{omega0}).

\subsection{Systematic weak-disorder expansion}

The above line of thought can be pursued
and yields a systematic weak-disorder expansion to all orders in the continuum
regime.

Let us look for a perturbative solution to~(\ref{feq}) of the form
\begin{equation}
F(y)=F_0(y)+F_2(y)+F_4(y)+\cdots,\qquad
\Omega=\Omega_0+\Omega_2+\Omega_4+\cdots
\end{equation}
The functions $F_{2k}(y)$ and the corresponding contributions $\Omega_{2k}$
to the weak-disorder expansion of the complex characteristic exponent
can be obtained recursively.

\subsubsection{Second order}

The equation for $F_2(y)$ reads
\begin{equation}
R_0(y)F_2(y)=-Q(y)F'_0(y)-R_2(y)F_0(y)+S_2(y)+2\Omega_2.
\end{equation}
The condition that $F_2(y)$ have no pole at the unstable zero ($y=z_-$)
yields
\begin{equation}
\Omega_2=\half\big[Q(y_-)F'_0(y_-)+R_2(z_-)F_0(z_-)-S_2(z_-)\big].
\end{equation}
We thus obtain the following explicit expression
for the second-order contribution $\Omega_2$:
\begin{eqnarray}
-8\mu^2\,\Omega_2
&=&(4\wbar^2+\ubar^2)\,\daa
\nonumber\\
&+&4\abar(\abar-\ubar)\,\dww
\nonumber\\
&+&\abar^2\,\duu
\nonumber\\
&+&4(-\ubar\mu-2\abar\,\wbar+\wbar\,\ubar)\,\daw
\nonumber\\
&+&2(2\wbar\mu-2\wbar^2-\abar\,\ubar)\,\dau
\nonumber\\
&+&4\abar(\wbar-\mu)\,\dwu.
\label{omega2}
\end{eqnarray}
We also obtain an explicit rational expression for $F_2(y)$,
with a triple pole at the stable zero ($y=z_+$).

It is worth pointing out that the result~(\ref{omega2}) agrees,
to leading order as the mean Iwasawa parameters $\abar$, $\wbar$ and $\ubar$
are simultaneously small,
with the following general expression
for the second-order weak-disorder expansion of the complex characteristic exponent
for arbitrary values of the mean parameters:
\begin{eqnarray}
-8\sin\abar\,\sinh^2\mu\,\Omega_2
&=&\sin\abar(\ubar^2\e^{2\wbar}+4\sinh^2\wbar)\,\daa
\nonumber\\
&+&4\sin^2\abar(\sin\abar-\ubar\cos\abar)\,\dww
\nonumber\\
&+&\sin^3\abar\,\e^{2\wbar}\,\duu
\nonumber\\
&-&8(\cosh(\wbar-\mu)-\cos\abar)(1-\cos\abar\,\e^{-\wbar-\mu})\,\daw
\nonumber\\
&-&4\sin\abar\,\e^{\wbar-\mu}(\cosh(\wbar-\mu)-\cos\abar)\,\dau
\nonumber\\
&-&4\sin^2\abar(\cos\abar-\e^{\wbar-\mu})\,\dwu.
\label{omega2gene}
\end{eqnarray}
The latter expression can be derived by means of
weak-disorder perturbative techniques
which have become standard in the case of $2\times2$ matrices~\cite{Lu}.
A systematic weak-disorder approach to the whole spectrum of Lyapunov exponents
of products of matrices of arbitrary order has been elaborated in~\cite{dmp}.

\subsubsection{Fourth order}

The equation for $F_4(y)$ reads
\begin{equation}
R_0(y)F_4(y)=-Q(y)F'_2(y)-R_2(y)F_2(y)+2\Omega_4.
\end{equation}
The condition that $F_4(y)$ have no pole at the unstable zero ($y=z_-$) yields
\begin{equation}
\Omega_4=\half\big[Q(z_-)F'_2(z_-)+R_2(z_-)F_2(z_-)\big].
\end{equation}
We thus obtain the lengthy expression
\begin{eqnarray}
-128\mu^5\Omega_4&=&(4\wbar^2+\ubar^2)(16\abar^2-16\abar\,\ubar+4\wbar^2+5\ubar^2)\daa^2
\nonumber\\
&+&16\abar(\abar-\ubar)(\abar^2-\abar\,\ubar+4\wbar^2)\dww^2
\nonumber\\
&+&5\abar^4\duu^2
\nonumber\\
&+&16\big[4\ubar\,\wbar(2\abar-\ubar)\mu+20\abar^2\wbar^2
\nonumber\\
&&\qquad+\abar^2\ubar^2-20\abar\,\wbar^2\ubar-\abar\,\ubar^3+4\wbar^2\ubar^2\big]\daw^2
\nonumber\\
&+&4\abar\big[-8\wbar(2\abar-\ubar)\mu+8\abar^2\ubar+20\abar\,\wbar^2-3\abar\,\ubar^2-8\ubar\,\wbar^2\big]\dau^2
\nonumber\\
&+&16\abar^2(-4\wbar\mu+\abar^2-\abar\,\ubar+4\wbar^2)\dwu^2
\nonumber\\
&+&8(20\abar^2\wbar^2-\abar^2\ubar^2+\abar\,\ubar^3-20\abar\,\ubar\,\wbar^2+6\ubar^2\wbar^2)\daa\dww
\nonumber\\
&+&2(8\abar^3\ubar+12\abar^2\wbar^2-3\abar^2\ubar^2+8\wbar^4)\daa\duu
\nonumber\\
&+&8(2\abar-\ubar)\big[-2\ubar(2\abar-\ubar)\mu
\nonumber\\
&&\qquad-\wbar(8\abar^2-8\abar\,\ubar+5\ubar^2+12\wbar^2)\big]\daa\daw
\nonumber\\
&+&4\big[4\wbar(2\abar-\ubar)^2\mu-8\abar^3\ubar-32\abar^2\wbar^2+4\abar^2\ubar^2-\abar\,\ubar^3
\nonumber\\
&&\qquad+20\abar\,\ubar\,\wbar^2-8\wbar^4-6\ubar^2\wbar^2\big]\daa\dau
\nonumber\\
&+&8\big[-2(2\abar-\ubar)(2\wbar^2+\abar\,\ubar)\mu
\nonumber\\
&&\qquad+\wbar(8\abar^3+12\abar\,\wbar^2-3\abar\,\ubar^2-8\ubar\,\wbar^2)\big]\daa\dwu
\nonumber\\
&+&8\abar^2(-\abar^2+\abar\,\ubar+6\wbar^2)\dww\duu
\nonumber\\
&+&32\wbar(-2\wbar\,\ubar\mu-6\abar^3+9\abar^2\ubar-4\abar\,\wbar^2-3\abar\,\ubar^2+2\ubar\,\wbar^2)\dww\daw
\nonumber\\
&+&16\abar\big[4(\abar-\ubar)\wbar\mu+\abar^2\ubar-10\abar\,\wbar^2-\abar\,\ubar^2+4\ubar\,\wbar^2\big]\dww\dau
\nonumber\\
&+&32\abar\,\wbar(-2\wbar\mu+3\abar^2-3\abar\,\ubar+2\wbar^2)\dww\dwu
\nonumber\\
&+&8\abar\big[2(-2\abar^2+\abar\,\ubar+2\wbar^2)\mu-\wbar(2\abar^2+3\abar\,\ubar+8\wbar^2)\big]\duu\daw
\nonumber\\
&+&4\abar^2(4\wbar\mu-4\abar^2-\abar\,\ubar-6\wbar^2)\duu\dau
\nonumber\\
&+&8\abar^3(5\wbar-2\mu)\duu\dwu
\nonumber\\
&+&16\big[2(2\abar-\ubar)(\abar\,\ubar-2\wbar^2)\mu
\nonumber\\
&&\qquad+\wbar(8\abar^3-4\abar^2\ubar+\abar\,\ubar^2+2\abar\,\wbar^2-4\ubar\,\wbar^2)\big]\daw\dau
\nonumber\\
&+&32(4\wbar^3\mu-4\abar^4+5\abar^3\ubar-2\abar^2\wbar^2-\abar^2\ubar^2-4\wbar^4)\daw\dwu
\nonumber\\
&+&16\abar\big[2(2\abar^2-\abar\,\ubar+2\wbar^2)\mu
\nonumber\\
&&\qquad+\wbar(-6\abar^2+\abar\,\ubar-4\wbar^2)\big]\dau\dwu.
\label{omega4}
\end{eqnarray}
We also get an explicit rational expression for $F_4(y)$,
with a pole of order 5 at the stable zero ($y=z_+$).

\subsubsection{Generic structure}

The structure of the weak-disorder expansion clearly appears
from the first two orders studied above.
At a generic (even) order $2k$ for $k\geq2$,
the equation for $F_{2k}(y)$ reads
\begin{equation}
R_0(y)F_{2k}(y)=-Q(y)F'_{2k-2}(y)-R_2(y)F_{2k-2}(y)+2\Omega_{2k}.
\end{equation}
The condition that $F_{2k}(y)$ have no pole at the unstable zero ($y=z_-$)
yields
\begin{equation}
\Omega_{2k}=\half\big[Q(z_-)F'_{2k-2}(z_-)+R_2(z_-)F_{2k-2}(z_-)\big].
\end{equation}
The structure of the expression thus obtained is clear from the recursive
nature
of the problem.
The denominator of $\Omega_{2k}$ is an integer multiple of $\mu^{3k-1}$.
The numerator is a homogeneous polynomial in the six covariances of degree $k$.
The explicit results obtained above at orders 2 and 4 suggest that the
\begin{equation}
d_k=\frac{(k+5)!}{k!\,5!}
\end{equation}
different possible monomials are all present in the end result.
We have $d_1=6$, $d_2=21$, $d_3=56$, $d_4=126$, and so on.
The coefficient of each monomial is a homogeneous polynomial
in $\abar$, $\wbar$, $\ubar$ and $\mu$, of degree $2k$ with integer
coefficients.
The degree in $\mu$ can be reduced to one by means of~(\ref{musca}).
Finally, $F_{2k}(y)$ is a rational function,
whose only pole is a multiple pole of order $2k+1$ at the stable zero
($y=z_+$).

\section{Monolithic disorder}
\label{sec:mono}

In this section we deal with the situations
where only one of the three variables $\a_n$, $w_n$ or $u_n$ is random.
We refer to these three special cases as {\em monolithic disorder}.

\subsection{Scalar disorder (only the $u_n$ are random)}
\label{sec:sca}

We began our study of this case in the introduction, where we derived
the following exact formula for the characteristic exponent
(see~(\ref{ressca}) and (\ref{gsca})):
\begin{equation}
\Omega=\b\,G(x),
\label{ressca2}
\end{equation}
where the scaling function $G(x)$ is
\begin{equation}
G(x)=\e^{-2\ii\pi/3}\frac{\Ai'(\e^{-2\ii\pi/3}x)}{\Ai(\e^{-2\ii\pi/3}x)}
=\frac{\Ai'(x)+\ii\,\Bi'(x)}{\Ai(x)+\ii\,\Bi(x)},
\label{gsca2}
\end{equation}
with
\begin{equation}
\b=\left(\frac{\abar^2\duu}{2}\right)^{1/3} \quad \text{and}\quad
x=\frac{\mu^2}{\b^2}.
\label{bbxdef}
\end{equation}

The argument~$x$ of the Airy functions involves $\mu$ and $\b$.
The first of these parameters is a measure of the distance to the band edge.
We have indeed $\mu^2<0$ inside the band,
i.e., on the support of the density of states in the absence of disorder,
whereas $\mu^2>0$ out of the band,
where the density of states vanishes in the absence of disorder.
The second parameter $\b$ entering~(\ref{bbxdef})
demonstrates that the effective disorder strength
is the product $\abar^2\duu$.
The precise correspondence with the quantum-mechanical problem is
discussed in Sec.~\ref{subsec:4.1QM}.

The Wronskian identity~\cite[(9.2.7), p.~194]{nist}, \cite[(10.4.10),
p.~446]{as}
\begin{equation}
\Ai'(x)\Bi(x)-\Ai(x)\Bi'(x)=-\frac{1}{\pi}
\end{equation}
yields in particular
\begin{equation}
\text{Im} \,G(x)=\frac{1}{\pi \left [ \Ai(x)^2+\Bi(x)^2 \right ]}.
\label{img}
\end{equation}

The above Airy scaling is characteristic of a generic band-edge singularity.
It has been met in a variety of situations,
including the problem of a white noise scalar
potential~\cite{FL,Ha,theo,sulem}
and the weak-disorder regime of the tight-binding Anderson model
with diagonal disorder near its band edges~\cite{dgedge}.

There remains to extract
from these formulae the behaviour of the characteristic exponent
in various interesting limits.
The differential equation obeyed
by the Airy functions~\cite[(9.2.1), p.~194]{nist}, \cite[(10.4.1), p.~446]{as}
\begin{equation}
\Ai''(x)-x\,\Ai(x)=0
\label{odesca}
\end{equation}
translates into the following Riccati equation
\begin{equation}
G(x)^2+G'(x)=x
\label{ricsca}
\end{equation}
for the scaling function $G(x)$.
Solving the above equation iteratively yields the asymptotic expansion
\begin{equation}
G(x) = x^{1/2}-\frac{1}{4x}-\frac{5}{32x^{5/2}}-\frac{15}{64x^4}+\cdots \quad
\text{as}\ x \rightarrow \infty,
\label{gexp}
\end{equation}
which corresponds to the weak-disorder expansion
\begin{equation}
\Omega = \mu-\frac{\abar^2\duu}{8\mu^2}-\frac{5\abar^4\duu^2}{128\mu^5}
-\frac{15\abar^6\duu^3}{512\mu^8}+\cdots
\quad\text{as}\ \duu \rightarrow 0.
\label{wdexp}
\end{equation}
The first two non-trivial terms are in agreement with~(\ref{omega2})
and~(\ref{omega4}).

It should be remarked, however,
that the expansion~(\ref{gexp}),
and therefore the weak-disorder expansion~(\ref{wdexp}),
are only divergent asymptotic expansions.
Indeed, these expansions fail to capture the exponentially small imaginary part of
the form
\begin{equation}
\text{Im}\, G(x)\sim\exp\left(-\frac{4}{3}x^{3/2}\right)
\label{nonpert}
\end{equation}
of the scaling function for large positive values of $x$, which
describes the tail of the density of states far away from the band,
i.e., deep into the region where there are no eigenstates in the absence of disorder.

We are led to draw from this explicit example the conclusion
that the weak-disorder expansion is only a divergent asymptotic expansion in general.
This observation should not be too much of a surprise,
if one remembers that perturbative techniques in other areas of physics,
including the well-known example of quantum field theory,
usually result in divergent asymptotic series.

Right at the band edge ($x=0$), we have
\begin{equation}
\Omega=\e^{-2\ii\pi/3}\frac{\Ai'(0)}{\Ai(0)}\,\b
=\e^{\ii\pi/3}\frac{\Gamma(2/3)}{\Gamma(1/3)}\left(\frac{3\abar^2\duu}{2}\right)^{1/3},
\end{equation}
and hence
\begin{equation}
\gamma=\frac{\Gamma(2/3)}{2\,\Gamma(1/3)}\left(\frac{3\abar^2\duu}{2}\right)^{1/3} \quad \text{and}\quad
j=\frac{\sqrt{3}\,\Gamma(2/3)}{2\pi\,\Gamma(1/3)}\left(\frac{3\abar^2\duu}{2}\right)^{1/3}.
\end{equation}

To close, let us consider the case where $\abar=\wbar=0$.
In this situation, the matrices
\begin{equation}
M_n=T_3(u_n)=\pmatrix{ 1 & u_n \cr 0 & 1 }
\end{equation}
belong to the nilpotent subgroup of $\text{SL}(2,{\mathbb R})$,
and hence commute among themselves.
We thus have $\Pi_n=T_3(U_n)$ with $U_n=u_1+\cdots+u_n\approx n\ubar$, and so
\begin{equation}
\Omega=0.
\label{omeganil}
\end{equation}
This result is recovered from~(\ref{ressca2}), since $\b=0$.

\subsection{Supersymmetric disorder (only the $w_n$ are random)}
\label{sec:susy}

We continue our study with the case where only the supersymmetric
variables~$w_n$ are
random.
The non-zero parameters are therefore $\abar$, $\wbar$, $\ubar$ and $\dww$.
The polynomial
\begin{equation}
Q(y)=4\dww y^2
\end{equation}
has a double zero at the origin and a double zero at infinity.
Equation~(\ref{heq}) yields
\begin{equation}
H(y)=y^{\nu-1}\,\e^{\Phi(y)},
\end{equation}
with
\begin{equation}
\nu=\frac{\wbar}{\dww}
\label{nudef}
\end{equation}
and
\begin{equation}
\Phi(y)=-\frac{1}{2\dww}\left(\abar y+\frac{\ubar-\abar}{y}\right).
\end{equation}
Equation~(\ref{keq}) reads
\begin{equation}
2\dww K'(y)=(\abar y-\wbar+\Omega)\,y^{-\nu-1}\,\e^{-\Phi(y)}.
\end{equation}
This function $K(y)$ must vanish both as $\abs{y}\to0$ and $\abs{y}\to\infty$
in the directions where $\Phi(y)$ diverges.
Assuming $\ubar>\abar>0$ for definiteness,
and changing the sign of the integration variable $y$ for convenience,
we obtain
\begin{equation}
\int_0^\infty(\abar y+\wbar-\Omega)\,y^{-\nu-1}\,\e^{\Phi(y)}\,\d y=0.
\end{equation}
By setting $t=1/y$,
the integrals involved in the latter condition can be reduced to integrals
of the form~\cite[vol.~I, (29), p.~146]{tit}
\begin{equation}
\int_0^\infty t^{\nu-1}\,\e^{-at-b/t}\,\d t
=2\left(\frac{b}{a}\right)^{\nu/2}\,K_\nu(2\sqrt{ab}),
\label{knudef}
\end{equation}
where $K_\nu$ is the modified Bessel function.
Using the identities~\cite[vol.~II, (25),~(26), p.~79]{htf}
\begin{equation}
x K_{\nu-1}(x)+\nu K_\nu(x)=x K_{\nu+1}(x)-\nu K_\nu(x)=-x K_\nu'(x),
\end{equation}
we eventually find
\begin{equation}
\Omega=\dww\,G(x)
\label{ressus}
\end{equation}
where
\begin{equation}
\fbox{$\displaystyle
G(x)=-x\,\frac{K'_\nu(x)}{K_\nu(x)}
$}
\label{gsus}
\end{equation}
and~(see~(\ref{nudef}))
\begin{equation}
\label{parsus}
\nu=\frac{\wbar}{\dww},\qquad x=\frac{\sqrt{\abar(\ubar-\abar)}}{\dww}.
\end{equation}
The characteristic exponent $\Omega$ is real and the stationary current $j$
vanishes
as long as the argument $x$ is real, i.e., $\abar(\ubar-\abar)>0$.

In the opposite situation, i.e., $\abar(\ubar-\abar)<0$,
it is more convenient to consider the real variable
\begin{equation}
\z=\frac{\sqrt{\abar(\abar-\ubar)}}{\dww}.
\end{equation}
The result~(\ref{ressus}) becomes
\begin{equation}
\Omega=\dww\,G_1(\z),
\label{ressus2}
\end{equation}
with
\begin{equation}
G_1(\z)=-\z\,\frac{J'_\nu(\z)-\ii N'_\nu(\z)}{J_\nu(\z)-\ii N_\nu(\z)},
\label{gsus2}
\end{equation}
where $J_\nu$ and $N_\nu$ are Bessel functions.
The Wronskian identity~\cite[vol.~II, (28), p.~79]{htf}
\begin{equation}
J'_\nu(\z)N_\nu(\z)-J_\nu(\z)N'_\nu(\z)=-\frac{2}{\pi\z}
\end{equation}
yields in particular
\begin{equation}
\text{Im}\,G_1(\z)=\frac{2}{\pi(J_\nu(\z)^2+N_\nu(\z)^2)}.
\label{ressusi}
\end{equation}

The above Bessel scaling functions have already been met in several
circumstances.
The forms~(\ref{gsus2}) and~(\ref{ressusi}) enter the analysis
of classical diffusion in a one-dimensional random force field~\cite{BCGL,bg},
whereas the form~(\ref{gsus}) shows up in
disordered supersymmetric quantum mechanics (see Sec.~\ref{subsec:4.2QM})
and in the quantum Ising chain in a disordered transverse magnetic
field~\cite{jmspin}.

The differential equation satisfied
by the modified Bessel function~\cite[vol.~II, (11), p.~5]{htf}
\begin{equation}
x^2K_\nu''(x)+xK_\nu'(x)-(x^2+\nu^2)K_\nu(x)=0
\label{kode}
\end{equation}
translates into the following Riccati equation
\begin{equation}
G(x)^2-xG'(x)=x^2+\nu^2=\frac{\mu^2}{\dww^2}
\label{ricsus}
\end{equation}
for the scaling function $G(x)$.
We use this equation to investigate the weak-disorder regime
where
\begin{equation}
\D(x)=\sqrt{x^2+\nu^2}=\frac{\mu}{\dww}
\end{equation}
is large. Solving~(\ref{ricsus}) iteratively yields the asymptotic expansion
\begin{equation}
G(x) = \D(x)+\frac{x^2}{2\D(x)^2}-\frac{x^2(x^2-4\nu^2)}{8\D(x)^5}+\cdots
\quad\text{as}\ \D(x) \rightarrow \infty.
\end{equation}
This corresponds to the weak-disorder expansion
\begin{equation}
\Omega = \mu+\frac{\abar(\ubar-\abar)\dww}{2\mu^2}
+\frac{\abar(\ubar-\abar)(\abar^2-\abar\,\ubar+4\wbar^2)\dww^2}{8\mu^5}
+\cdots
\label{fwsus}
\end{equation}
as $\dww \rightarrow 0$, in agreement with~(\ref{omega2}) and~(\ref{omega4}).

In the opposite strong-disorder regime,
the expansion at small $x$
\begin{equation}
G(x)=\abs{\nu}+\frac{x^2}{2(1+\abs{\nu})}+\cdots
\label{gnuexpand}
\end{equation}
translates into
\begin{equation}
\Omega=\abs{\wbar}+\frac{\abar(\ubar-\abar)}{2\dww}+\cdots
\end{equation}

Let us turn to the case $\wbar=0$, where the index $\nu$ vanishes.
This situation, which will also be met for distance disorder with
$\ubar=2\abar$,
corresponds to a critical point.
In the present case we have
\begin{equation}
x=\frac{\mu}{\dww},
\end{equation}
while the scaling law~(\ref{ressus}) becomes
\begin{equation}
\Omega=\mu\,\frac{K_1(x)}{K_0(x)}.
\end{equation}
The weak-disorder expansion~(\ref{fwsus}) simplifies to
\begin{equation}
\Omega = \mu+\frac{\dww}{2}-\frac{\dww^2}{8\mu}+\cdots
\end{equation}
In the opposite regime of a strong disorder, corresponding to $x\to0$,
the logarithmic singularity $K_0(x)\approx\ln(2/x)-\C$
translates into the singular behavior
\begin{equation}
\Omega\approx\frac{\dww}{\ln(2\dww/\mu)-\C} \quad\text{as}\ \dww \rightarrow
\infty,
\label{lnww}
\end{equation}
where $\C$ denotes Euler's constant.

To close, let us consider the case where $\abar=\ubar=0$.
In this situation, the matrices
\begin{equation}
M_n=T_2(w_n)=\pmatrix{\e^{w_n} & 0 \cr 0 & \e^{-w_n} }
\end{equation}
belong to the Abelian (diagonal) subgroup of $\text{SL}(2,{\mathbb R})$,
and hence commute among themselves.
We thus have $\Pi_n=T_2(W_n)$ with $W_n=w_1+\cdots+w_n\approx n\wbar$, and so
\begin{equation}
\Omega=\abs{\wbar}.
\label{omegaabe}
\end{equation}
This result is recovered from~(\ref{gnuexpand}), since $x=0$.

\subsection{Distance disorder (only the $\a_n$ are random)}
\label{sec:dis}

We close this section with the case where only the distance variables $\a_n$
are random.
The non-zero parameters are therefore $\abar$, $\wbar$, $\ubar$ and $\daa$.
The polynomial
\begin{equation}
Q(y)=\daa(y^2+1)^2
\end{equation}
has two double zeros at $y=\pm\ii$.
Equation~(\ref{heq}) yields
\begin{equation}
H(y)=
\left(\frac{1-\ii y}{1+\ii y}\right)^{\ii\lambda}
\frac{\e^{\Phi(y)}}{y^2+1}.
\end{equation}
with
\begin{equation}
\lambda=\frac{\ubar-2\abar}{2\daa},\qquad
\Phi(y)=\frac{\ubar y-2\wbar}{\daa(y^2+1)}.
\label{defdis}
\end{equation}
Equation~(\ref{keq}) reads
\begin{equation}
\daa K'(y)=
(S(y)+2\Omega)
\left(\frac{1+\ii y}{1-\ii y}\right)^{\ii\lambda}
\frac{\e^{-\Phi(y)}}{y^2+1},
\end{equation}
where $S(y)$ is given by~(\ref{pols}).

The analysis is facilitated by working with the variable
\begin{equation}
t=\frac{1+\ii y}{1-\ii y}
\end{equation}
which maps the lower half-plane ($\text{Im} \,y<0$)
onto the complement of the unit circle ($\abs{t}>1$),
and the double zeros ($y=\ii$ and $y=-\ii$)
to zero and infinity respectively.
The transformed function $\hK(t) := K(y)$ satisfies the differential equation
\begin{eqnarray}
\ii\daa\hK'(t)
&=&\left [ (\Omega-\wbar-\ii\abar)t
+(\Omega-\wbar+\ii\abar)\frac{1}{t}+2(\Omega-\wbar+\daa)\right ]
\nonumber\\
&\times&\frac{t^{\ii\lambda}}{(t+1)^2}\,\e^{-\hPhi(t)},
\label{dcdtdis}
\end{eqnarray}
with $\hPhi(t) := \Phi(y)$, i.e.,
\begin{equation}
\hPhi(t)
=-\frac{1}{4\daa}\left [
(2\wbar+\ii\ubar)t+(2\wbar-\ii\ubar)\frac{1}{t}+4\wbar\right ].
\end{equation}

Let us assume $\wbar<0$ for definiteness.
The function $\hK(t)$ must vanish as $\text{Re}\, t\to -\infty$.
The characteristic exponent $\Omega$ is therefore determined by the condition
\begin{equation}
\int_{C}\hK'(t)\,\d t=0,
\end{equation}
where the integration contour $C$ circles around the branch cut
of the integrand along the negative real axis.

An integration by parts, corresponding to the choice
\begin{equation}
g(y)=-\frac{y}{y^2+1}
\label{fdis}
\end{equation}
of the gauge function
(this trick will be explained in detail in Sec.~\ref{sec:gene};
see Equation~(\ref{sfdef})),
simplifies the above equation to
\begin{equation}
\int_{C}
\left [ (2\wbar+\ii\ubar)t+(2\wbar-\ii\ubar)\frac{1}{t}-4\Omega\right ]
t^{\ii\lambda-1}\,\e^{-\hPhi(t)}\,\d t=0.
\label{intdis}
\end{equation}
The integrals involved in the latter expression are of the form
\begin{equation}
\int_{C} t^{\nu-1}\,\e^{at+b/t}\,\d t
=2\pi\ii\left(\frac{b}{a}\right)^{\nu/2}\!\!I_\nu(2\sqrt{ab}),
\label{inudef}
\end{equation}
where $I_\nu$ is the modified Bessel function.
This result, which bears a close resemblance with~(\ref{knudef}),
can be easily derived from~\cite[(8.412.2), p.~954]{gr}.
Using the identities~\cite[vol.~II, (23), (24), p.~79]{htf}
\begin{equation}
x I_{\nu+1}(x)+\nu I_\nu(x)=x I_{\nu-1}(x)-\nu I_\nu(x)=x I_\nu'(x),
\end{equation}
we eventually obtain
\begin{equation}
\Omega=\daa\,G(x),
\label{resdis}
\end{equation}
where
\begin{equation}
\fbox{$\displaystyle
G(x)=x\,\frac{I'_{\ii\lambda}(x)}{I_{\ii\lambda}(x)},
$}
\label{gdis}
\end{equation}
and~(see~(\ref{defdis}))
\begin{equation}
\lambda=\frac{\ubar-2\abar}{2\daa},\qquad
x=\frac{\sqrt{4\wbar^2+\ubar^2}}{2\daa}.
\label{pardis}
\end{equation}
The present case is in some sense dual to the previous one.
Both scaling functions~(\ref{gsus}) and~(\ref{gdis}) involve a modified Bessel
function.
In the present situation of distance disorder
the argument $x$ is real and the index $\nu=\ii\lambda$ is imaginary,
whereas in the previous case of supersymmetric disorder (see~(\ref{parsus})),
$x$ can be either real or imaginary and the index $\nu$ is real.

The differential equation~(\ref{kode}) is also satisfied by $I_\nu(x)$. We can
study the weak-disorder regime by
using the corresponding Riccati equation for the scaling function $G(x)$:
\begin{equation}
G(x)^2+xG'(x)=x^2-\lambda^2=\frac{\mu^2}{\daa^2}.
\label{ricdis}
\end{equation}
Set
\begin{equation}
\D(x)=\sqrt{x^2-\lambda^2}=\frac{\mu}{\daa}.
\end{equation}
By solving~(\ref{ricdis}) iteratively, we obtain the asymptotic expansion
\begin{equation}
G(x) = \D(x)-\frac{x^2}{2\D(x)^2}-\frac{x^2(x^2+4\lambda^2)}{8\D(x)^5}+\cdots
\quad\text{as}\ x\rightarrow \infty.
\end{equation}
This corresponds to the weak-disorder expansion
\begin{eqnarray}
\Omega = \mu&-&\frac{(4\wbar^2+\ubar^2)\daa}{8\mu^2}
\nonumber\\
&-&\frac{(4\wbar^2+\ubar^2)(4\wbar^2+5\ubar^2-16\abar\,\ubar+16\abar^2)\daa^2}{128\mu^5}
+\cdots
\label{fwdis}
\end{eqnarray}
as $\daa \rightarrow 0$, in agreement with~(\ref{omega2}) and~(\ref{omega4}).

In the strong-disorder regime, the small-$x$ expansion
\begin{equation}
G(x) = -\ii\lambda+\frac{x^2}{2(1-\ii\lambda)}+\cdots
\label{glambdaexpand}
\end{equation}
translates into
\begin{equation}
\Omega=\ii\,\frac{2\abar-\ubar}{2}+\frac{4\wbar^2+\ubar^2}{8\daa}+\cdots
\quad\text{as}\ \daa \rightarrow \infty.
\label{wstrong}
\end{equation}

Let us turn to the case where $\ubar=2\abar$, so that the index $\lambda$
vanishes.
In this case we have
\begin{equation}
x=\frac{\mu}{\daa},
\end{equation}
while the scaling law~(\ref{resdis}) becomes
\begin{equation}
\Omega=\mu\,\frac{I_1(x)}{I_0(x)}.
\end{equation}
The weak-disorder expansion~(\ref{fwdis}) simplifies to
\begin{equation}
\Omega=\mu-\frac{\daa}{2}-\frac{\daa^2}{8\mu}+\cdots \quad\text{as}\ \daa
\rightarrow 0.
\end{equation}
In the opposite regime of a strong disorder, corresponding to $x\to0$,
we obtain the estimate
\begin{equation}
\Omega\approx\frac{\mu^2}{2\daa} \quad\text{as}\ \daa \rightarrow \infty.
\end{equation}

To close, let us consider the case where $\wbar=\ubar=0$.
In this situation, the matrices
\begin{equation}
M_n=T_1(\a_n)=\pmatrix{\cos\a_n&-\sin\a_n\cr\sin\a_n&\cos\a_n}
\end{equation}
belong to the compact (rotation) subgroup of $\text{SL}(2,{\mathbb R})$,
and hence commute among themselves.
We thus have $\Pi_n=T_1(A_n)$ with $A_n=\a_1+\cdots+\a_n\approx n\abar$, and so
$\gamma=0$,
while $j=\pm\abar/\pi$.
The sign of $j$ depends on how the limiting case is reached.
Our prescription $\wbar\to 0^-$ yields
\begin{equation}
\Omega=\ii\pi j=\ii\abar.
\label{omegacom}
\end{equation}
This result is recovered from~(\ref{glambdaexpand}), since $x=0$.

\section{General potential disorder}
\label{sec:pot}

In this section we deal with the case of a general potential disorder,
whose supersymmetric and scalar parts $\dw_n$ and $\du_n$ are correlated.
The non-zero parameters are therefore $\abar$, $\wbar$, $\ubar$
and the three covariances $\dww$, $\dwu$ and~$\duu$.
For further convenience we introduce the reduced quantities
\begin{equation}
c=\frac{\dwu}{\sqrt{\dww\duu}},\qquad b=\sqrt{1-c^2},\qquad
y_0=\half\sqrt{\frac{\duu}{\dww}}.
\label{cbydef}
\end{equation}
The polynomial $Q(y)$ then reads
\begin{equation}
Q(y)=4\dww(y^2+2cy_0y+y_0^2),
\end{equation}
where $c$ is nothing but the correlation coefficient between $\dw$ and $\du$.

\subsection{The fully correlated cases ($c^2=1$)}

Let us consider the fully correlated cases where $c=\pm1$.
The polynomial
\begin{equation}
Q(y)=4\dww(y+cy_0)^2
\end{equation}
has a double zero on the real axis at $y=-cy_0$
and a double zero at infinity.
Equation~(\ref{heq}) yields
\begin{equation}
H(y)=(y+cy_0)^{\nu-1}\,\e^{\Phi(y)},
\end{equation}
where
\begin{equation}
\Phi(y)=-\frac{1}{2\dww}\left(\abar y+\frac{\b}{y+cy_0}\right),
\end{equation}
and
\begin{equation}
\nu=\frac{\wbar+\abar cy_0}{\dww},\qquad
\b=\ubar-\abar+2(\dww-\wbar)cy_0-\abar y_0^2.
\label{nubdef}
\end{equation}
Equation~(\ref{keq}) reads
\begin{equation}
2\dww K'(y)=(\abar y-\wbar+\Omega)(y+cy_0)^{-\nu-1}\,\e^{-\Phi(y)}.
\end{equation}
Assuming $\b>0$ and $\abar>0$ for definiteness,
the function $K(y)$ must vanish both as $y\to -cy_0-\ii 0$ and as
$y\to-\infty$,
where $\Phi(y)$ diverges.
Setting $t=-1/(y+cy_0)$,
the integrals involved in the above condition can again be performed
in terms of modified Bessel functions.
Our end result reads
\begin{equation}
\Omega=\dww\,G(x),
\label{restot}
\end{equation}
where the scaling function
\begin{equation}
\label{eq:ggpdfc}
G(x)=-x\,\frac{K'_\nu(x)}{K_\nu(x)}
\end{equation}
is identical to that of the supersymmetric monolithic
disorder~(see~(\ref{gsus}));
we will give an explanation of this observation at the end of
Sec.~\ref{sec:MixedSchrodinger}.
Its argument reads $x=\sqrt{\abar\b}/\dww$, i.e.,
\begin{equation}
x=\frac{\sqrt{\abar\left [ \ubar-\abar+2(\dww-\wbar)cy_0-\abar
y_0^2\right ]}}{\dww}.
\end{equation}
We recall that $c=\pm1$,
while $y_0$ and $\nu$ have been defined in~(\ref{cbydef}) and~(\ref{nubdef}).

\subsection{The partly correlated case ($c^2<1$)}

Let us now turn to the generic partly correlated case where $c^2<1$, so that
$b>0$.
The polynomial
\begin{equation}
Q(y)=4\dww(y^2+2cy_0y+y_0^2)
\end{equation}
has two simple zeros at the complex conjugate points
\begin{equation}
y_1=(-c-\ii b)y_0,\qquad
y_2=(-c+\ii b)y_0,
\end{equation}
and a double zero at infinity.
Equation~(\ref{heq}) yields
\begin{equation}
H(y)=(y-y_1)^{\nu_1-1}(y-y_2)^{\nu_2-1}\,\e^{\Phi(y)},
\end{equation}
where
\begin{eqnarray}
\nu_1&=&\frac{b+\ii c}{2b}+
\frac{\ii(\ubar-\abar)+2\wbar(b-\ii c)y_0+\ii\abar(b-\ii
c)^2y_0^2}{4\dww by_0},\nonumber\\
\nu_2&=&\frac{b-\ii c}{2b}+
\frac{\ii(\abar-\ubar)+2\wbar(b+\ii c)y_0-\ii\abar(b+\ii c)^2y_0^2}{4\dww by_0}
\end{eqnarray}
and
\begin{equation}
\Phi(y)=-\frac{\abar y}{2\dww}.
\end{equation}
Equation~(\ref{keq}) reads
\begin{equation}
2\dww K'(y)=(\abar
y-\wbar+\Omega)(y-y_1)^{-\nu_1}(y-y_2)^{-\nu_2}\,\e^{-\Phi(y)}.
\end{equation}
Assuming $\text{Re} \,\nu_1<1$ for definiteness,
the function $K(y)$ must vanish both as $y\to y_1-\ii 0$ and as $y\to-\infty$.
The change of variable
\begin{equation}
t=\frac{y-y_1}{y_1-y_2},
\end{equation}
mapping $y_1$ onto 0 and $y_2$ onto $-1$,
leads us to the condition
\begin{eqnarray}
\int_0^\infty \Bigl \{ \left [ \dww\,x(2t+1)+\dww(\nu_1+\nu_2-1)-\Omega\right ]
\nonumber\\
\times\, t^{-\nu_1}(t+1)^{-\nu_2}\,\e^{-xt} \Bigr\} \,\d t=0,
\end{eqnarray}
with
\begin{equation}
x=\frac{\ii\abar by_0}{\dww}=\frac{\ii\abar\sqrt{\dww\duu-\dwu^2}}{2\dww^2}.
\label{xpartdef}
\end{equation}
The integrals involved in the above condition can be expressed in terms
of the Whittaker function~\cite[vol.~I, (18), p.~274]{htf}, \cite[(9.220),
p.~1059]{gr}
\begin{equation}
W_{l,m}(x)=\frac{x^{m+\half}\,\e^{-\half x}}{\Gamma(m-l+\half)}
\int_0^\infty t^{m-l-\half}(t+1)^{m+l+\half}\,\e^{-xt}\,\d t
\end{equation}
and of its derivative with respect to~$x$.
The end result is
\begin{equation}
\Omega=\dww\,G(x),
\label{respart}
\end{equation}
where the argument $x$ has been defined in~(\ref{xpartdef}),
and the scaling function $G(x)$ reads
\begin{equation}
\fbox{$\displaystyle
G(x)=1-2x\frac{W'_{l,m}(x)}{W_{l,m}(x)}.
$}
\label{gwhi}
\end{equation}
In this expression,
\begin{eqnarray}
\label{eq:ParameterL}
l&=&\half(\nu_1-\nu_2)
=\frac{\ii c}{2b}+ \ii \frac{\ubar-\abar-2\wbar cy_0+\abar(1-2c^2)y_0^2}{4\dww
by_0},
\\
\label{eq:ParameterM}
m&=&\half(1-\nu_1-\nu_2)=-\frac{\wbar+\abar cy_0}{2\dww}
=-\frac1{2\dww}\left(\wbar+\frac{\abar\dwu}{2\dww}\right),
\end{eqnarray}
so that $l$ is imaginary and $m$ real.
The numbers $c$, $b$ and $y_0$ were defined in~(\ref{cbydef}).

In Sec.~\ref{sec:MixedSchrodinger},
we will establish the relation between the result (\ref{gwhi}) and a
result of~\cite{HagTex08} for a continuous model: the
Schr\"odinger equation with mixed disorder.

Interestingly, a characteristic function with a
similar structure ---although not exactly the same--- appears when
considering
a product of elements in the two-parameter subgroup of
$\text{SL}(2,{\mathbb R})$ obtained by setting $w=0$ in the Iwasawa
decomposition:
\begin{equation}
M= \pmatrix{\cos \a & -\sin \a \cr
\sin \a & \cos \a}\, \pmatrix{1 & u \cr
0 & 1}.
\end{equation}
When $\a$ and $u$ are two independent exponential variables,
the matrices
$M_n$ are not close to the identity matrix, and yet it is possible to obtain
the invariant measure $f$ and
the characteristic function $\Omega$ in closed analytical form.
Such a product of matrices corresponds to a quantum-mechanical model for
$\delta$-impurities such that the spacing between consecutive impurities and
the strength of each impurity are independent and exponentially distributed
(see
Sec.~\ref{subsec:QM1DPointLikeScatterers}).
This model was first solved by Nieuwenhuizen~\cite{theo}
and studied again in a broader context in~\cite{CTT1}.

The differential equation satisfied
by the Whittaker function~\cite[(9.222), p.~1060]{gr}
\begin{equation}
4x^2W_{l,m}''(x)-(x^2-4lx+4m^2-1)W_{l,m}(x)=0
\end{equation}
translates into the following Riccati equation:
\begin{equation}
G(x)^2-2xG'(x)=x^2-4lx+4m^2=\frac{\mu^2+\abar\dwu}{\dww^2}
\label{ricwhi}
\end{equation}
for the scaling function $G(x)$.
Setting
\begin{equation}
\D(x)=\sqrt{x^2-4lx+4m^2}=\frac{\sqrt{\mu^2+\abar\dwu}}{\dww},
\end{equation}
and solving~(\ref{ricwhi}) iteratively yields the large-$\D(x)$ asymptotic
expansion
\begin{equation}
G(x) = \D(x)+\frac{x(x-2l)}{\D(x)^2}-\frac{x \left [ x^3+4(l^2-4m^2)x+16lm^2
\right ]}{2\D(x)^5}+\cdots
\end{equation}
When this is expressed in terms of the covariances,
we recover the weak-disorder expansion results~(\ref{omega2})
and~(\ref{omega4}).

\section{Independent disorder with zero mean}
\label{sec:ell}

In this section we deal with the case where the three random variables
$\a_n$, $w_n$ and $u_n$ are independent and have zero mean.
The non-zero parameters are therefore the three variances $\daa$, $\dww$ and
$\duu$.
In this situation the matrices $M_n$ fluctuate around the unit matrix according
to
\begin{equation}
M_n-I \approx\pmatrix{w_n&u_n-\a_n\cr\a_n&-w_n}.
\end{equation}
The mean square norm of the difference between $M_n$ and the unit matrix reads
\begin{equation}
\eps^2=\mean(\norm{M_n-I}^2)=2\daa+2\dww+\duu.
\label{eps2}
\end{equation}

The polynomial $Q(y)$ reads
\begin{equation}
Q(y)=\daa y^4+2(\daa+2\dww)y^2+\daa+\duu.
\end{equation}
Let us henceforth consider the generic situation where none of the variances
vanishes.
This is the first instance where $Q(y)$ has four distinct zeros.
Furthermore we have the polynomial identities
\begin{equation}
\sigma(z)\,\sigma'(z)=2 v(z) \quad\text{and}\quad
Q'(y)=2R(y).
\label{elliden}
\end{equation}

The first of the above identities ensures that the condition~(\ref{j0})
is automatically fulfilled.
The stationary current $j$ therefore vanishes,
and the characteristic exponent $\Omega$ is real.

The second of the identities~(\ref{elliden}) yields
\begin{equation}
H(y)=\frac{1}{\sqrt{Q(y)}}.
\end{equation}
Equation~(\ref{keq}) then reads
\begin{equation}
K'(y)=\frac{\daa(y^2+1)+2\Omega}{\sqrt{Q(y)}}.
\end{equation}

The next step consists of investigating the zeros of $Q(y)$.
First, $Q(y)$ is an even polynomial.
Setting $t=y^2$, we have
\begin{equation}
Q(y)=\daa t^2+2(\daa+2\dww)t+\daa+\duu.
\end{equation}
The two zeros of this expression read
\begin{eqnarray}
t_1&=&-\frac{\daa+2\dww+\sqrt{(\daa+2\dww)^2-\D^2}}{\daa},\nonumber\\
t_2&=&-\frac{\daa+2\dww-\sqrt{(\daa+2\dww)^2-\D^2}}{\daa},
\label{ellt}
\end{eqnarray}
with
\begin{equation}
\D=\sqrt{\daa(\daa+\duu)}.
\label{elldel}
\end{equation}
Two regimes are to be considered separately.

\subsubsection*{Regime~I ($\D<\daa+2\dww$, i.e., $\duu<4\dww(\daa+\dww)/\daa$)}

In this first regime, $t_1$ and $t_2$ are real and obey $t_1<t_2<0$.
The four zeros of $Q(y)$ are ordered as follows along the imaginary axis:
\begin{equation}
y_1=-\ii\sqrt{-t_1},\qquad
y_2=-\ii\sqrt{-t_2},\qquad
y_3=\ii\sqrt{-t_2},\qquad
y_4=\ii\sqrt{-t_1}.
\label{elly}
\end{equation}

The function $K(y)$ must vanish at both zeros of $Q(y)$ which lie in the lower
half-plane,
i.e., $y_1$ and $y_2$, where $H(y)$ diverges.
We thus obtain the condition
\begin{equation}
\int_{y_1}^{y_2}\frac{\daa(y^2+1)+2\Omega}{\sqrt{Q(y)}}\,\d y=0.
\label{ellint}
\end{equation}
An integration by parts, corresponding to the choice
\begin{equation}
g(y)=-\frac{1}{y-y_4}
\label{fell}
\end{equation}
of the gauge function (see~(\ref{sfdef})),
removes the term in $y^2$ in the numerator of the integrand.
The change of variable
\begin{equation}
y=\frac{y_1(y_4-y_2)+y_4(y_2-y_1)\sin^2\theta}{y_4-y_2+(y_2-y_1)\sin^2\theta}
\label{ellch}
\end{equation}
allows one to express the remaining integrals in terms of the Legendre
complete elliptic integrals
\begin{equation}
\K(k)=\int_0^{\pi/2}\frac{\d\theta}{\sqrt{1-k^2\sin^2\theta}},\qquad
\E(k)=\int_0^{\pi/2}\sqrt{1-k^2\sin^2\theta}\,\d\theta,
\end{equation}
where the square elliptic modulus is the cross ratio
\begin{equation}
k^2=(y_1,y_4;y_2,y_3):=\frac{(y_1-y_2)(y_4-y_3)}{(y_1-y_3)(y_4-y_2)}.
\label{ellcross}
\end{equation}

The zeros $y_1,\dots,y_4$ can be eliminated from the resulting expression
by means of~(\ref{ellt}) and~(\ref{elly}).
Some algebra leads us to the following end result
\begin{equation}
\Omega=\half\left((\daa+2\dww+\D)\frac{\E(k)}{\K(k)}-(\daa+\D)\right),
\label{ellr1}
\end{equation}
where the elliptic modulus reads
\begin{equation}
k=\sqrt{\frac{\daa+2\dww-\D}{\daa+2\dww+\D}},
\label{ellk}
\end{equation}
with $\D$ being defined in~(\ref{elldel}).

Finally, the differential identity~\cite[vol.~II, p.~322]{htf},
\cite[(8.123.2), p.~907]{gr}
\begin{equation}
\E(k)=(1-k^2)(\K(k)+k\K'(k))
\end{equation}
and the linear differential equation~\cite[(8.124.1), p.~907]{gr}
\begin{equation}
k(k^2-1)\K''(k)+(3k^2-1)\K'(k)+k\K(k)=0
\end{equation}
allow one to recast the above result as
\begin{equation}
\Omega=\D\, G(k)+\dww,
\label{gell}
\end{equation}
where the scaling function
\begin{equation}
\fbox{$\displaystyle
G(k)=k\frac{\K'(k)}{\K(k)}-\frac{k^2}{1-k^2}
$}
\end{equation}
obeys the Riccati equation
\begin{equation}
G(k)^2+kG'(k)=-\frac{k^2}{(1-k^2)^2}=\frac{\daa\duu-4\dww(\daa+\dww)}{4\daa(\daa+\duu)}.
\label{ricell1}
\end{equation}

\subsubsection*{Regime~II ($\D>\daa+2\dww$, i.e.,
$\duu>4\dww(\daa+\dww)/\daa$)}

In this second regime, $t_1$ and $t_2$ form a complex conjugate pair.
The four zeros of~$Q(y)$ are still given by~(\ref{elly}).
They now sit at the vertices of a rectangle in the complex plane,
with $y_1$ and $y_2$ being in the lower half-plane,
and their respective opposites $y_4$ and $y_3$ in the upper half-plane.
The elliptic modulus~$k$ is accordingly found to be purely imaginary.
The characteristic exponent $\Omega$ can be obtained
as the analytical continuation of the result~(\ref{ellr1}) to imaginary values
of $k$.
It is however advantageous to perform the change of modulus~\cite[vol.~II,
Ch.~13]{htf}
from $k^2$ to
\begin{equation}
\kk^2=-\frac{k^2}{1-k^2}=(y_2,y_4;y_1,y_3):=\frac{(y_2-y_1)(y_4-y_3)}{(y_2-y_3)(y_4-y_1)}.
\label{ellcross2}
\end{equation}
Our result~(\ref{ellr1}) thus becomes
\begin{equation}
\Omega=\D\,\frac{\E(\kk)}{\K(\kk)}-\half(\daa+\D),
\label{ellr2}
\end{equation}
with
\begin{equation}
\kk=\sqrt{\frac{\D-\daa-2\dww}{2\D}}.
\label{ellk2}
\end{equation}
The above result can be recast as
\begin{equation}
\Omega=\D\,\GG(\kk)+\dww,
\label{gell2}
\end{equation}
where the scaling function $\GG(\kk)\equiv G(k)$, i.e.,
\begin{equation}
\fbox{$\displaystyle
\GG(\kk)=\kk(1-\kk^2)\frac{\K'(\kk)}{\K(\kk)},
$}
\end{equation}
obeys the Riccati equation
\begin{equation}
\GG(\kk)^2+\kk(1-\kk^2)\GG'(\kk)=\kk^2(1-\kk^2)
=\frac{\daa\duu-4\dww(\daa+\dww)}{4\daa(\daa+\duu)}.
\label{ricell2}
\end{equation}
The rightmost side of~(\ref{ricell1}) and~(\ref{ricell2})
is the same rational expression of the variances,
which is negative in the first case and positive in the second.

In the borderline situation where $\duu=4\dww(\daa+\dww)/\daa$,
so that $\D=\daa+2\dww$ and $k=\kk=0$, all the above expressions consistently
yield
\begin{equation}
\Omega=\dww.
\end{equation}

The above general results~(\ref{ellr1}),~(\ref{ellr2})
can be made more explicit in the special cases where one of the variances
vanishes.

\subsection{Supersymmetric~and scalar disorder ($\daa=0$)}

In this case,~(\ref{ellr1}) leads to the simple result
\begin{equation}
\Omega=0.
\end{equation}
Indeed $\daa=0$ yields $\D=0$ and $k=1$; the result therefore follows from
the fact that $\E(1)=1$, whilst $\K(k)$ diverges to infinity as
\begin{equation}
\K(k) \approx \ln\frac{4}{\sqrt{1-k^2}}
\label{klog}
\end{equation}
in the $k\to1$ limit~\cite[(8.113.3),~p.~905]{gr}.

\subsection{Distance~and supersymmetric disorder ($\duu=0$)}

In this case, we have $\D=\daa$, and so the results~(\ref{ellr1}),~(\ref{ellk})
become
\begin{equation}
\Omega=(\daa+\dww)\frac{\E(k)}{\K(k)}-\daa \quad\text{and}\quad
k=\sqrt{\frac{\dww}{\daa+\dww}}.
\end{equation}
Further simplifications arise in various regimes.

For $\dww\ll\daa$, i.e., $k\to0$, the
expansions~\cite[(8.113.3),~p.~905]{gr}
\begin{equation}
\K(k)=\frac{\pi}{2}\sum_{n\ge0}a_nk^{2n},\quad
\E(k)=\frac{\pi}{2}\sum_{n\ge0}a_n\frac{k^{2n}}{1-2n},\quad
a_n=\frac{(2n)!^2}{(2^nn!)^4}
\label{ellser}
\end{equation}
of the elliptic integrals yield
\begin{equation}
\Omega = \frac{\dww}{2}-\frac{\dww^2}{16\daa}+\frac{\dww^3}{32\daa^2}+\cdots
\end{equation}

In the opposite regime ($\dww\gg\daa$), the behaviour~(\ref{klog}) yields
\begin{equation}
\Omega \approx \frac{2\dww}{\ln(16\dww/\daa)}.
\end{equation}
This inverse logarithmic singularity is reminiscent of~(\ref{lnww}).

When both variances are equal ($\dww=\daa$), the modulus reads $k=1/\sqrt2$.
In this case, the elliptic integrals can be expressed
in terms of Gauss's lemniscate constant
\begin{equation}
\G=\frac{\Gamma(1/4)^2}{(2\pi)^{3/2}}=0.834626\dots
\label{GLC}
\end{equation}
We thus end up with
\begin{equation}
\Omega=\frac{\daa}{\pi\G^2}=0.456946\dots\daa.
\end{equation}

\subsection{Distance~and scalar disorder ($\dww=0$)}

In this case, the results~(\ref{ellr2}),~(\ref{ellk2}) read
\begin{equation}
\Omega=\D\,\frac{\E(\kk)}{\K(\kk)}-\half(\daa+\D),\qquad
\kk=\sqrt{\frac{\D-\daa}{2\D}}.
\end{equation}
Further simplifications occur in certain limits.

For $\duu\ll\daa$, the series~(\ref{ellser}) for the elliptic
integrals yield
\begin{equation}
\Omega =
\frac{\duu}{8}-\frac{9\duu^2}{256\daa}+\frac{39\duu^3}{2048\daa^2}+\cdots
\end{equation}

In the opposite regime ($\duu\gg\daa$),
the elliptic moduli go to $k=\ii$ and $\kk=1/\sqrt2$, so that we get
\begin{equation}
\Omega\approx\frac{\sqrt{\daa\duu}}{2\pi\G^2}=0.228473\dots\sqrt{\daa\duu}.
\end{equation}

\subsection{Maximal Lyapunov exponent}

The complex characteristic exponent $\Omega$
has been found to vanish whenever two of the variances vanish.
This property is due to the fact that the matrices $M_n$
have a trivial composition law in these circumstances:
they commute among themselves.
The above property can thus be viewed as a special case
of~(\ref{omeganil}),~(\ref{omegaabe}),~(\ref{omegacom}).

It is therefore of interest to look at the maximum of the characteristic
exponent,
for a fixed value of the mean square norm $\eps^2$ (see~(\ref{eps2})).
A numerical investigation of the result~(\ref{ellr1})
shows that $\Omega$ reaches its maximum
\begin{equation}
\Omega=0.170787995\dots\eps^2
\end{equation}
for the following values of the variances:
$\daa=0.042658\dots\eps^2$,
$\dww=0.416226\dots\eps^2$,
$\duu=0.082228\dots\eps^2$.
The corresponding elliptic modulus reads $k=0.919798\dots$

\section{The general case}
\label{sec:gene}

We now turn to the general case, where the nine parameters of the problem,
namely the three mean variables and the six covariances, take generic values.

\subsection{Preliminaries}

The polynomial $Q(y)$ generically has four distinct zeros $y_i$
($i=1,\dots,4$).
It can therefore be factored as
\begin{equation}
Q(y)=\daa\prod_{i=1}^4(y-y_i).
\end{equation}
It will also prove useful to introduce the partial fraction expansion
\begin{equation}
\frac{R(y)}{Q(y)}=\sum_{i=1}^4\frac{a_i}{y-y_i},
\label{rqratio}
\end{equation}
where the exponents
\begin{equation}
a_i=\frac{R(y_i)}{Q'(y_i)}.
\end{equation}
are, in general, complex numbers.
Expanding~(\ref{rqratio}) around $\infty$ to order $1/y$, we obtain the sum
rule
\begin{equation}
\sum_{i=1}^4a_i=2.
\label{sum}
\end{equation}
Equation~(\ref{heq}) yields
\begin{equation}
H(y)=\prod_{i=1}^4(y-y_i)^{-a_i},
\end{equation}
so that~(\ref{keq}) reads
\begin{equation}
\daa K'(y)=(S(y)+2\Omega)\prod_{i=1}^4(y-y_i)^{a_i-1}.
\end{equation}

The polynomial $Q(y)=\sigma^2(y)$ is manifestly positive
for all real values of~$y$.
As a consequence, no zero of $Q(y)$ can cross the real line.
Under generic circumstances,
two zeros of $Q(y)$, say $y_1$ and $y_2$, are in the lower half-plane,
whereas the other two are in the upper half-plane.
We choose to label those zeros so as to have
\begin{equation}
y_3=y_2^*,\quad y_4=y_1^*,\qquad
a_3=a_2^*,\quad a_4=a_1^*.
\label{choice}
\end{equation}

\subsection{Calculation of the characteristic exponent}

Let us consider for definiteness
the case where the exponents $a_1$ and $a_2$ have positive real parts.
In this situation, the function $K(y)$ must vanish as $y=y_1$ and $y=y_2$,
where~$H(y)$ diverges.
We thus obtain the condition
\begin{equation}
\int_{y_1}^{y_2}(S(y)+2\Omega)\prod_{i=1}^4(y-y_i)^{a_i-1}\,\d y=0.
\label{ogene1}
\end{equation}

The above expression has a form of gauge invariance.
This property has already been mentioned in two special cases
in Sec.~\ref{sec:dis} and~\ref{sec:ell}.
It is now time to explain it in full generality.
Consider an arbitrary regular function $g(y)$.
Equation~(\ref{heq}) implies
\begin{equation}
\frac{\d}{\d y}\,\frac{g(y)}{H(y)}=\frac{Q(y)g'(y)+R(y)g(y)}{Q(y)H(y)},
\end{equation}
and so
\begin{equation}
\int_{y_1}^{y_2}\frac{Q(y)g'(y)+R(y)g(y)}{Q(y)H(y)}\,\d y=0.
\end{equation}
As a consequence, the expression~(\ref{ogene1})
for the characteristic exponent $\Omega$ is left invariant if $S(y)$ is
replaced
by
\begin{equation}
S^{[g]}(y)=S(y)+Q(y)g'(y)+R(y)g(y).
\label{sfdef}
\end{equation}
The integrations by parts performed so far to simplify the expression for
$\Omega$
in two special cases in Sec.~\ref{sec:dis} and~\ref{sec:ell}
correspond to simple rational choices for the gauge function~$g(y)$
(see~(\ref{fdis}),~(\ref{fell})).

In the general case under consideration,
it is advantageous to consider the gauge function
\begin{equation}
g(y)=-\frac{1}{y-y_2},
\label{fgene}
\end{equation}
where $y_2$ is the endpoint of the integral~(\ref{ogene1}).
We then have
\begin{equation}
S^{[g]}(y)=\frac{A(y_2)}{y-y_2}+B(y_2),
\end{equation}
where
\begin{equation}
A(y_2)=Q'(y_2)-R(y_2),\qquad
B(y_2)=\half Q''(y_2)-R'(y_2)+S(y_2)
\end{equation}
are polynomials in $y_2$ of respective degrees 3 and 2.
The condition~(\ref{ogene1}) therefore becomes
\begin{equation}
\int_{y_1}^{y_2}\left(\frac{A(y_2)}{y-y_2}+B(y_2)+2\Omega\right)
\prod_{i=1}^4(y-y_i)^{a_i-1}\,\d y=0.
\label{ogene2}
\end{equation}

Now, let us perform the rational change of variable
\begin{equation}
t=\frac{y-y_1}{y_2-y},
\end{equation}
which maps $y_1$ to 0, $y_2$ to $\infty$, $\infty$ to $-1$,
and $y_4$ and $y_3$ to $-1/\ka$ and $-1/\la$ respectively, where
\begin{equation}
\ka=\frac{y_4-y_2}{y_4-y_1},\qquad\la=\frac{y_3-y_2}{y_3-y_1}.
\end{equation}
The integrals involved in~(\ref{ogene2}) are thus reduced to integrals
of the form~\cite[vol.~I, (35), p.~312]{tit}
\begin{eqnarray}
\int_0^\infty\!(1+\ka t)^{a-1}(1+\la t)^{b-1}t^{c-1}\d t
&=&\frac{\Gamma(c)\Gamma(2-a-b-c)}{\ka^c\,\Gamma(2-a-b)}
\nonumber\\
&\times&\F(1-b,c;2-a-b;x),
\label{igene}
\end{eqnarray}
where the argument
\begin{equation}
x=1-\frac{\la}{\ka}
=(y_1,y_4;y_2,y_3):=\frac{(y_1-y_2)(y_4-y_3)}{(y_1-y_3)(y_4-y_2)}
\label{xdef}
\end{equation}
is the same cross ratio of the four zeros
as the square elliptic modulus $k^2$ of~(\ref{ellcross}),
while
\begin{equation}
\F(\a,\b;\g;x)=\frac{\Gamma(\g)}{\Gamma(\a)\Gamma(\b)}
\sum_{n\ge0}\frac{\Gamma(n+\a)\Gamma(n+\b)}{\Gamma(n+\g)}\,\frac{x^n}{n!}
\end{equation}
is Gauss's hypergeometric function.

We thus get the following expression
\begin{eqnarray}
\Omega=\half\Biggl [\!&\daa&(y_1-y_4)(y_3-y_2)\,a_1\,
\frac{\F(1-a_3,a_1+1;a_1+a_2;x)}{\F(1-a_3,a_1;a_1+a_2;x)}\nonumber\\
+&\daa&(y_2-y_3)(y_2-y_4)(1-a_2)-B(y_2)\Biggr ].
\label{ogene3}
\end{eqnarray}

Finally, the differential relation~\cite[vol.~I, (23), p.~102]{htf}
\begin{equation}
\b\,\F(\a,\b+1;\g;x)=\b\,\F(\a,\b;\g;x)+x\,\F'(\a,\b;\g;x)
\end{equation}
allows one to bring the above result to the compact scaling form
\begin{equation}
\Omega=\half\left(\daa(y_1-y_4)(y_3-y_2)\,G(x)
+\frac{R(y_1)}{y_1-y_3}-S(y_1)\right),
\label{ghyp}
\end{equation}
where $x$ is the cross ratio introduced in~(\ref{xdef}),
and the scaling function $G(x)$ reads
\begin{equation}
\fbox{$\displaystyle
G(x)=x\,\frac{\F'(1-a_3,a_1;a_1+a_2;x)}{\F(1-a_3,a_1;a_1+a_2;x)}.
$}
\label{Ghyp}
\end{equation}

Our end result~(\ref{ghyp}) involves,
besides basic parameters,
the complex zeros $y_i$ of the polynomial $Q(y)$
and the associated complex exponents $a_i$.
The labeling~(\ref{choice}) of the complex zeros
allows us to recast the argument $x$ as
\begin{equation}
x=\abs{\frac{y_2-y_1}{y_2^*-y_1}}^2.
\end{equation}
This quantity manifestly obeys $0<x<1$.
This condition, together with
\begin{equation}
\text{Re} \left (a_1 + a_2 \right ) = 1,
\end{equation}
ensures that the hypergeometric series entering~(\ref{Ghyp}) are convergent.
The above result therefore provides a meaningful expression
of the complex characteristic exponent for generic parameter values.

The differential equation satisfied by Gauss's hypergeometric function
\begin{eqnarray}
x(1-x)\F''(\a,\b;\g;x)&+&(\g-(\a+\b+1)x)\F'(\a,\b;\g;x)
\nonumber\\
&-&\a\b\,\F(\a,\b;\g;x)=0
\end{eqnarray}
translates into the following Riccati equation
\begin{eqnarray}
(1-x)G(x)^2&+& x(1-x)G'(x)
\nonumber\\
&+&\left [ (a_3-a_1-1)x+a_1+a_2-1 \right ]G(x) \nonumber \\
&=&a_1(1-a_3)x\,
\label{richyp}
\end{eqnarray}
for the scaling function $G(x)$.
Equation~(\ref{richyp}) provides a useful check of the weak-disorder regime,
where all the exponents $a_i$ become simultaneously large.
To leading order, we obtain
\begin{equation}
G(x)=\frac{2(\abar y_1y_3-\wbar(y_1+y_3)-\ubar+\abar\pm\mu(y_1-y_3))}
{\daa(y_1-y_3)(y_1-y_4)(y_2-y_3)}.
\end{equation}
Choosing the lower sign, we recover after some algebra the expected result
$\Omega=\mu$.

It is worth emphasizing that both the sum rule~(\ref{sum})
and the choice~(\ref{fgene}) of the gauge function
are essential for the above reduction to be worked out.
For instance, the integrals involved in the original expression~(\ref{ogene1}),
with $S(y)$ being a quadratic polynomial,
would involve an extra multiple pole at $t=-1$,
and could therefore not be reduced to the form~(\ref{igene}).
Other admissible choices of the gauge function $g(y)$
and/or of the rational change of variable from $y$ to~$t$
correspond to permuting the complex zeros $y_i$.
The effect of these discrete transformations is to change the argument $x$
into one of the six possible values of the cross ratio of the four zeros:
\begin{equation}
\left\{
x,\quad
1-x,\quad\frac{1}{x},\quad\frac{1}{1-x},\quad\frac{x}{x-1},\quad\frac{x-1}{x}
\right\}.
\end{equation}
Only the first two possibilities ensure that the argument is in the range
$0<x<1$.

\subsection{The case of a vanishing stationary current}
\label{vanishing}

To close, we return to the case where the stationary probability current
(integrated density of states, rotation number) $j$ vanishes,
so that $\Omega$ is real.

The condition for this to happen has been derived in
Sec.~\ref{invariantSubsection}.
Using~(\ref{qrres}), Equation~(\ref{j0}) can be recast as
\begin{equation}
\int_{-\infty}^\infty \frac{R(z)}{Q(z)}\,\d z=0.
\end{equation}
Finally, the partial fraction expansion~(\ref{rqratio}) allows one
to perform the above integral explicitly, along the lines of~(\ref{log}).
The result is $\ii\pi(a_3+a_4-a_1-a_2)$.
Using the relation~(\ref{sum}), we are thus left with the condition
\begin{equation}
a_1+a_2=1.
\label{cond}
\end{equation}
In other words, as a consequence of~(\ref{sum}) and~(\ref{choice}),
the exponents $a_1$ and $a_2$ obey in general
\begin{equation}
\text{Re}\,(a_1+a_2)=1,
\label{gal}
\end{equation}
while the condition~(\ref{j0}) for the stationary current
$j$ to vanish is equivalent to the extra condition
\begin{equation}
\text{Im}\,(a_1+a_2)=0.
\end{equation}
The resulting expression for the scaling function,
\begin{equation}
\fbox{$\displaystyle
G(x)=x\,\frac{\F'(a_1,a_1^*;1;x)}{\F(a_1,a_1^*;1;x)},
$}
\end{equation}
is then manifestly real.
It can be checked that the full expression~(\ref{ghyp})
for the characteristic exponent is also real in this case.

The condition~(\ref{cond}) can be expanded in terms of the three mean Iwasawa
parameters
and of the six covariances by eliminating the zeros $y_i$.
We thus obtain a formidable homogeneous polynomial equation of degree six
which consists of 335 terms.

If we restrict the analysis to the situation where the mean parameters
$\abar$, $\wbar$ and $\ubar$ vanish,
the condition~(\ref{cond}) reduces to a homogeneous polynomial equation
in the covariances with only 22 terms.
Its explicit form
\begin{eqnarray}
&&32\daw^4\dwu^2+16\daa\daw^2\dau\dwu^2-32\dww\daw^3\dau\dwu
\nonumber\\
&-&8\daa(\daa+\dww)\daw\dwu^3-8\daa(\daa+2\dww)\daw^2\dwu^2
\nonumber\\
&+&2\daa^2\dau^2\dwu^2+8\dww\duu\daw^3\dwu-8\daa\dww\dau^2\daw\dwu
\nonumber\\
&+&2\duu^2\daw^4+8\dww^2\dau^2\daw^2-2\daa^2\dwu^4
\nonumber\\
&+&(4\dww^2+4\daa\dww-\daa\duu)
\nonumber\\
&\times&(\daa\dwu^2+2\daa\daw\dwu-\duu\daw^2)\dau=0
\end{eqnarray}
is however not very illuminating.

To close, let us mention that the above condition does not directly apply
to the particular cases studied so far,
as they correspond to multiple zeros,
whereas it was implicitly assumed in the derivation of~(\ref{cond})
that the four zeros of $Q(y)$ were simple.

For independent disorder with zero mean (Sec.~\ref{sec:ell}),
we have shown that $j$ vanishes identically.
This case is very special, as the polynomial identities~(\ref{elliden})
imply $a_i=1/2$ for $i=1,\dots,4$.

\subsection{Relationship to the work of Zanon and Derrida}
\label{sec:zd}

As already underlined in the introduction,
the present work can be viewed as a systematic treatment
of the degenerate weak-disorder expansion put forward by Zanon and Derrida~\cite{zd}.
These authors observe that the systematic perturbative expansion
of Reference~\cite{dmp} breaks down when the unperturbed matrix has degenerate eigenvalues,
and that the simplest case where this occurs
corresponds to $2\times2$ matrices close to the identity,
i.e., essentially the subject of this work.
More explicitly, they investigate the behavior of the Lyapunov exponent
of products of real matrices
$A_n\in\text{GL}(2,{\mathbb R})$
of the form
\begin{equation}
A_n=I+\pmatrix{a_n&b_n\cr c_n&d_n},
\end{equation}
in the weak-disorder regime
where the four random variables $a_n,\dots,d_n$ are simultaneously small,
having zero mean values ($\ave{a}=\dots=\ave{d}=0$)
and arbitrary covariances $\ave{a^2}$, $\ave{ab}$, and so on.

The connection between the above parametrization of the matrices $A_n$
and the Iwasawa parametrization of matrices
$M_n\in\text{SL}(2,{\mathbb R})$
is made by setting
\begin{equation}
A_n=M_n\sqrt{\Delta_n},
\end{equation}
with
\begin{equation}
\Delta_n=\mathrm{det}A_n=(1+a_n)(1+d_n)-b_nc_n,
\end{equation}
and using~(\ref{GS}).
We thus obtain
\begin{eqnarray}
\a_n&=&\arctan\frac{c_n}{1+a_n},\nonumber\\
w_n&=&\frac12\ln\frac{(1+a_n)^2+c_n^2}{(1+a_n)(1+d_n)-b_nc_n},\nonumber\\
u_n&=&\frac{(1+a_n)b_n+c_n(1+d_n)}{(1+a_n)^2+c_n^2},
\end{eqnarray}
whereas the respective characteristic exponents $\Gamma$ and $\Omega$
of the products of matrices $A_n$ and $M_n$ are related by
\begin{equation}
\Gamma=\Omega+\frac12\,\ave{\ln\Delta}.
\label{GO}
\end{equation}

In the weak-disorder regime defined above, we have therefore
\begin{eqnarray}
\abar&=&-\ave{ac},
\quad
\wbar=\frac14(-\ave{a^2}+2\ave{c^2}+\ave{d^2}+2\ave{bc}),
\quad
\ubar=-\ave{ab}-2\ave{ac}+\ave{cd},
\nonumber\\
\daa&=&\ave{c^2},
\quad
\dww=\frac14(\ave{a^2}+\ave{d^2}-2\ave{ad}),
\quad
\duu=\ave{b^2}+\ave{c^2}+2\ave{bc},
\\
\daw&=&\frac12(\ave{ac}-\ave{cd}),
\quad
\dau=\ave{bc}+\ave{c^2},
\quad
\dwu=\frac12(\ave{ab}+\ave{ac}-\ave{bd}-\ave{cd}),
\nonumber
\end{eqnarray}
whereas~(\ref{GO}) translates into
\begin{equation}
\Gamma=\Omega-\frac14(\ave{a^2}+\ave{d^2}+2\ave{bc}).
\end{equation}

Our general result~(\ref{ghyp}) therefore yields
an exact scaling formula for the characteristic exponent $\Gamma$
in terms of the ten covariances $\ave{a^2},\ave{ab},\dots$
Such a formula was not derived in~\cite{zd};
the key novel ingredient allowing for a substantial progress
in the present work consists in the use of the Hilbert transform.
The present approach also allows one to easily recover the explicit
results by Zanon and Derrida in the following two examples.

\begin{itemize}
\item[$\bullet$] {\it Example~1}.
The non-diagonal elements $b_n$ and $c_n$ are i.i.d. variables,
and so the non-zero covariances are $\ave{b^2}=\ave{c^2}=\sigma^2$.
In the language of this work, the non-zero parameters are
$\wbar=\sigma^2/2$, $\daa=\dau=\sigma^2$ and $\duu=2\sigma^2$.
The polynomial $Q(y)=\sigma^2(y^4+1)$ has four simple zeros at the vertices of a square,
and we have $a_1=a_2=1/2$.
This is therefore an elliptic case,
albeit not of the canonical form studied in Sec.~\ref{sec:ell}.
Our result~(\ref{ghyp}) yields
\begin{equation}
\Gamma=\frac{\sigma^2}{2\pi\G^2}=0.228473\dots\sigma^2,
\end{equation}
where $\G$ is Gauss's lemniscate constant (see~(\ref{GLC})),
in agreement with~\cite{zd} (up to a factor 2 in the numerical value).
\item[$\bullet$] {\it Example~2}.
The four elements $a_n,\dots,d_n$ are i.i.d. variables,
and so the non-zero covariances are $\ave{a^2}=\ave{b^2}=\ave{c^2}=\ave{d^2}=\sigma^2$.
In the language of this work, the non-zero parameters are
$\wbar=\dww=\sigma^2/2$, $\daa=\dau=\sigma^2$ and $\duu=2\sigma^2$.
The polynomial $Q(y)=\sigma^2(y^2+1)^2$ has double zeros at $y=\pm\ii$.
This is a very special case of the monolithic distance disorder considered in Sec.~\ref{sec:dis}.
Equation~(\ref{keq}) reads $K'(y)=1+2\Gamma/(\sigma^2(y^2+1))$.
This equation has non-integrable singularities at the zeros, unless $\Gamma=0$.
The latter result was obtained in~\cite{zd}.
\end{itemize}

\section{A limiting case: Hyperbolic Brownian motion}
\label{sec:HyperbolicBM}

In the previous sections we have shown how to compute the characteristic
exponent of an infinite product of random matrices in the continuum regime
where the matrices are close to the identity.
This approach provides a rather complete understanding of one-dimensional
soluble random potentials. The knowledge of the Lyapunov exponent brings
valuable information on the degree of localization of the wave functions of an
infinite disordered sample. Another observable, probably more suited for finite
samples, is the reflexion phase acquired by a particle which is scattered on
the boundary of the sample.
In the limit of a semi-infinite system, this phase converges to a
random variable with a non-trivial distribution which has been studied
by several authors~\cite{sulem}. In particular Barnes and Luck~\cite{BaLu}
have shown how to relate it to the invariant measure of
the Riccati variable.

In this section we will return to this problem in the case of general point
scatterers. This problem can be reformulated in geometric terms as a random
walk on the hyperbolic plane. For certain choices of parameters of our model,
this random walk converges in the continuum regime to a hyperbolic Brownian
motion~\cite{FJ}.
The purpose of this section is two-fold. On the one hand we demonstrate by
an explicit calculation that hyperbolic Brownian motion is indeed
included in our general scheme. Moreover we will use this
correspondence to study the convergence of the Lyapunov exponent to
its bulk value as a function of the size of the system.

We consider a disordered sample of finite length $L$. We assume that the
potential which consists of $n$ point scatterers has its support on the
interval $[0,L]$. In this setting we now consider the following scattering
problem
\begin{itemize}
\item[$\bullet$]
For $x\leq 0$, the wave function
$\psi(x)=\e^{-\ii kx} + r(k)\, \e^{\ii kx}$ represents an incoming
wave which is partially reflected by the sample.
\item[$\bullet$]
For $x\geq L$, the wave function is a transmitted wave $\psi(x)= t(k)\,\e^{\ii
kx}$.
\end{itemize}
We have
\begin{equation}
\pmatrix{ \psi' (0) \cr \psi (0) }
= \Pi_n^{-1} \pmatrix{ \psi' (L) \cr \psi (L) },
\end{equation}
where
\begin{equation}
\Pi_n^{-1} := M_{1}^{-1} M_{2}^{-1} \cdots M_{n}^{-1}
\label{productOfMatrices2}
\end{equation}
and the $M_n$ are given by~(\ref{waveFunction}).
Note that the corresponding Riccati variable may be expressed in terms
of the scattering data
\begin{equation}
\frac{z(0)}{k}=\frac{\psi'(0)}{k\,\psi(0)}=\ii\,\frac{1-r(k)}{1+r(k)}
\quad \mbox{and} \quad
\frac{z(L)}{k}=\frac{\psi'(L)}{k\,\psi(L)}=\ii.
\end{equation}
Set $k=1$ for definiteness.
The scattering problem by point scatterers can be
reformulated as an iteration of Moebius maps
\begin{equation}
{\mathcal M}^{-1} (z) = \frac{m_{22} z - m_{12}}{m_{11}-m_{21} z}
\end{equation}
(see~(\ref{mtfdef}) and the notations set below~(\ref{mdef})),
acting on the upper half-plane of the complex variable $z=x+\ii y$.\\
The Riccati variable $z(0)$ can be expressed in terms of the reversed iterates
of the point $z=\ii$
\begin{equation}
z(0)= \widehat{z}_n(\ii)
:=
\left(
{\mathcal M}^{-1}_1\circ{\mathcal M}^{-1}_2\circ \cdots\circ
{\mathcal M}^{-1}_n
\right) (\ii).
\end{equation}
The backward sequence $ \left\{ \widehat{z}_n(\ii) \right\} $ can be
viewed as the successive approximants of a random continued fraction
which converges to a random limit for $n\rightarrow \infty$.
Standard results~\cite{ChLe} show that
it converges to a real random variable $x= \lim_{n\to\infty}
\widehat{z}_n(\ii)$ whose density is given by the invariant measure
$f(x)$.

In physical terms, this expresses the fact that, for a system of large
size, the incoming wave is entirely reflected ($r=\e^{\ii\delta}$),
implying $z(0)=\cot(\delta/2)$.\\
The random orbit $ \left\{ \widehat{z}_n(\ii) \right\} $ defines a
random walk on the complex plane which can be conveniently described
in geometric terms using standard tools of hyperbolic geometry which
we briefly recall.

\subsection{The Poincar\'e upper half-plane}

Defined as
\begin{equation}
{\mathbb H}= \left\{z=x+\ii y \:\vert\: x,\,y\in {\mathbb R}, y\geq0\right\},
\end{equation}
it is globally invariant under the action of
\begin{equation}
M := \pmatrix{ m_{11} & m_{12} \cr m_{21} & m_{22} }
\in \text{SL}(2,{\mathbb R}).
\end{equation}
The hyperbolic distance between two points $z,\,z'\in {\mathbb H} $, given by
\begin{equation}
\cosh d(z,z')=\frac{(x-x')^2+y^2+y'^2}{2yy'},
\label{distance}
\end{equation}
is invariant under the group action:
\begin{equation}
d(\mathcal{M}(z),\mathcal{M}(z'))=d(z,z')
\quad \forall \;M \in \text{SL}(2,{\mathbb R}).
\label{group invariance}
\end{equation}
Using the identity
\begin{equation}
2\, \cosh d(\ii,\mathcal{M}(\ii))= |M|^2
\end{equation}
and the group invariance (\ref{group invariance}) one can compute the distance
between two successive iterates of the random walk,
\begin{equation}
\cosh d(\widehat{z}_n(\ii),\widehat{z}_{n-1}(\ii))
=\cosh d(\ii,\mathcal{M}_n(\ii))
=\frac12|M_n|^2,
\end{equation}
which shows that the step length distribution only depends on the
subgroup corresponding to
\begin{equation}
\pmatrix{\e^w & 0 \cr
0 & \e^{-w}} \,
\pmatrix{1 & u \cr
0 & 1}
\end{equation}
in the Iwasawa decomposition.

\subsection{Continuum limit}

When the matrices $M_n$ are close to the identity, the recursion relation
for $T_n=\Pi_n^{-1}$, i.e.,
\begin{equation}
T_{n+1} -T_{n} = T_{n}\, (M^{-1}_{n+1}-I),
\label{recursion}
\end{equation}
turns into a stochastic differential equation {\em on the group}
$\text{SL}(2,{\mathbb R})$ whose
general form we proceed to discuss.

To this end, it is helpful to work with the {\em finite-dimensional}
representation of
$\text{SL}(2,{\mathbb R})$ which associates to an element $M$ the operator
corresponding to
matrix multiplication by $M$. Introduce the infinitesimal generators $X_i$, $i
\in \{\a,w,u\}$, defined implicitly by
\begin{equation}
\e^{\a X_a} = \pmatrix{\cos \a & -\sin \a \cr \sin \a & \cos \a},\quad
\e^{w X_w} = \pmatrix{\e^w & 0 \cr 0 & \e^{-w}},\quad
\e^{u X_u} = \pmatrix{1 & u \cr 0 & 1}.
\end{equation}
One can show that
$(M^{-1}_{n+1}-I)$ will converge to a Lie algebra-valued
white noise with drift.
Therefore, replacing the index $n$ by the continuous
variable~$t$, the recurrence (\ref{recursion}) goes into
\begin{equation}
\deriv{ T_t }{t} = T_t
\sum_{i \in \{\alpha,\,w,\,u\}}
\left [ \overline{\lambda}_i X_i + X_i\,\eta_i(t) \right ],
\label{equation stochastique}
\end{equation}
where the $\eta_i(t)$ are three white noises with some
correlation matrix and the~$\ave\lambda_i$ are drift terms expressible in
terms of the means $\abar$, $\wbar$, $\ubar$.

In the following we will analyse a particular case and show how it
fits into our general scheme.
We assume that the non-zero parameters of the model are $\dww=1/4$,
$\duu=1$, $\ave\lambda_w=\wbar=-\eps/2$.
Therefore
\begin{equation}
\sum_{i \in \{\a,w,u\}} ( \ave\lambda_i X_i + X_i \eta_i(t) )
= -\frac{\eps}{2}X_w
+ \frac{1}{2}X_w\eta_w(t) +X_u\eta_u(t).
\end{equation}
In our case, the Iwasawa decomposition of $T_t$ is of the form
\begin{equation}
\label {matrix T}
T_t=\pmatrix{1 & x(t)\cr 0 & 1}\pmatrix{\sqrt {y(t)} & 0\cr 0&
\frac{1}{\sqrt{y(t)}} }.
\end{equation}
The action on ${z=\ii}$ yields the following process in ${\mathbb H}$:
\begin{equation}
z(t)=\mathcal{T}_t (\ii)= x(t)+\ii\, y(t).
\end{equation}
The Stratonovich differential equation on $\text{SL}(2,{\mathbb R})$
(\ref{equation
stochastique}) is then lifted to a Stratonovich differential
equation on ${\mathbb H}$:
\begin{equation}
\dot x(t)=y(t)\,\eta_u(t)
\quad
\mbox{and}
\quad
\dot y(t)=-\eps\, y(t)+y(t)\,\eta_w(t).
\end{equation}
By integration we obtain
\begin{equation}
y(t)=\e^{-\eps t+B(t)},
\quad
\mbox{where}
\quad
B(t) :=\int_0^t\d t'\,\eta_w(t').
\end{equation}
The process $y(t)$ is therefore simply the exponential of a linear
Brownian motion with drift.
In a suitable clock $\tau(t)$ the process $x(t)$ is also a Brownian
motion~\cite{CM,MY}:
introducing the time transformation
$t\to\tau(t)$ defined by
\begin{equation}
\deriv{\tau(t)}{t}=y(t)^2,
\quad\mbox{i.e.,}\quad
\tau(t)= \int_0^{t} \d t'\, \e^{-2\eps t'+2B(t') },
\label {exponential}
\end{equation}
and using
$\eta_u(\tau(t))\eqlaw\eta_u(t)/\sqrt{\deriv{\tau(t)}{t}}$, we get
\begin{equation}
\deriv{x(t)}{\tau(t)}=\eta_u(\tau(t)),
\end{equation}
which gives
\begin{equation}
\label{eq:822}
x(t)=\int_0^{\tau(t)}\d \tau'\, \eta_u(\tau').
\end{equation}
Using these results we can compute the Lyapunov exponent
\begin{equation}
\gamma=\lim_{t\to\infty}\frac{\ln|T_t|}{t},
\end{equation}
as well as
the invariant measure $f(x)$ and show that they do agree with the results obtained from
the general formulae (\ref{ifp}) and (\ref{simplifiedLyapunov}). In fact, since we know
explicitly the joint law $ \left\{ x(t),\, y(t) \right\} $, we can even
discuss the case of a finite system and obtain the convergence to the
semi-infinite system in terms of a central limit theorem.
Using (\ref{matrix T}) gives
\begin{equation}
\lim_{t\to\infty}\frac{\ln|T_t|-\eps\,t/2}{\sqrt{t}}
\eqlaw\frac{1}{2} {\tt N}(0,1),
\end{equation}
where ${\tt N}(0,1)$ is the standard normal variable.
The Lyapunov exponent is therefore $\gamma=\eps/2$.
The convergence to a normal variable is in agreement with standard results
on products of random matrices~\cite{BL}.
Note however that the variance differs from that given
by single-parameter scaling theory (SPS)~\cite{AndThoAbrFis80}.
The present example thus provides a novel explicit case where SPS
is not satisfied~\cite{CohRotSha88,deych,SchTit03,schrader}.

The invariant measure may be derived from the limiting law of (\ref{eq:822}).
The exponential functional of Brownian motion (\ref{exponential}) converges to
a random variable
on $(0,\,\infty)$ with probability density
\begin{equation}
\varrho (\tau) = \frac{2}{\Gamma(\eps)} \left (\frac{1}{2\tau}\right
)^{1+\eps}\e^{-\frac{1}{2\tau}}.
\end{equation}
Therefore
\begin{equation}
f(x)= \int_0^{\infty}\d\tau\,\frac{1}{\sqrt{2\pi\tau}}
\e^{-\frac{x^2}{2\tau}} \varrho(\tau)
=\frac{\Gamma(\eps+1/2)}{\sqrt{\pi}\Gamma(\eps) (x^2+1)^{\eps+1/2}}.
\end{equation}
One can check that this is indeed a zero-current solution of
(\ref{ifp}) for the choice of parameters $\dww=1/4$, $\duu=1$,
$\wbar=-\eps/2$.

For $\eps=1/2$ the invariant measure is given by the well-known Poisson
kernel which is known to be the exit law of Brownian motion on
${\mathbb H}$.

\section{Disordered quantum-mechanical interpretation}
\label{sec:MixedSchrodinger}

\subsection{One-dimensional quantum mechanics with point-like scatterers}
\label{subsec:QM1DPointLikeScatterers}

As mentioned in the introduction and discussed at length in~\cite{CTT1}, any
product of $2 \times 2$ matrices can be mapped to a Schr\"odinger equation
of the form
\begin{equation}
H \psi(x) = E \, \psi(x)
\quad
\mbox{with}
\quad
E=k^2,
\end{equation}
where the Hamiltonian involves a potential combining two random functions
\begin{equation}
\label{eq:HamiltonianMixed}
H = -\deriv{^2}{x^2} + W_1(x)^2 - W_1'(x) + W_2(x).
\end{equation}
The free Hamiltonian
\begin{equation}
H_0=-\deriv{^2}{x^2}
\end{equation}
is related to the
compact component of the Iwasawa decomposition, i.e., to the rotation matrix
\begin{equation}
\pmatrix{\cos \a & -\sin \a \cr
\sin \a & \cos \a}.
\end{equation}
On the other hand, $W_1$ and $W_2$ are related to the
Abelian and nilpotent components, i.e., to the diagonal and upper triangular
matrices
\begin{equation}
\pmatrix{\e^w & 0 \cr
0 & \e^{-w}} \quad\text{and}\quad
\pmatrix{1 & u \cr
0 & 1},
\end{equation}
respectively.
As Equation~(\ref{eq:TransferMatrixForPsi}) makes clear,
the product (\ref{productOfMatrices}) plays the r\^{o}le of a transfer
matrix acting on the wave function and its derivative.
The precise correspondence between the product of matrices and the
quantum-mechanical model is obtained by letting
$\ell_n:=\a_n/k$ be
the distance between two consecutive scatterers,
so that the position of the $n$th scatterer is given by
$x_n:=\sum_{p=1}^{n}\ell_p$. (Of course, this only makes sense if $\a_n > 0$.)
Set
\begin{eqnarray}
\label{eq:DefW1}
&& W_1(x) = \lim_{\varepsilon \rightarrow 0+} \sum_n w_n \, \delta( x -
x_n+\varepsilon ),\\
\label{eq:DefW2}
&& W_2(x) = \lim_{\varepsilon \rightarrow 0+} \sum_n v_n \, \delta( x -
x_n-\varepsilon ),
\end{eqnarray}
where $v_n = k u_n$ and $\varepsilon\to0^+$.
Then the Hamiltonian (\ref{eq:HamiltonianMixed}) describes an
irregular lattice of double impurities, in the terminology of~\cite{CTT1}.
The case where the impurity strength $u_n$ is distributed according to an
exponential law
and $w_n=0$ was considered in~\cite{CTT1,theo}, while the converse problem
where $u_n=0$ and $w_n$
is exponentially distributed was solved in~\cite{CTT1}.

\subsection{Continuum limit for $\daa=0$}

In this paper, we have considered the limit of small random
parameters $\a_n$, $u_n$ and $w_n$,
retaining only the statistical information contained in the first two cumulants.
Within this scheme, the constraint $\a_n\geq0$ can only be satisfied by setting
$\daa=0$, so that the distance
$\ell_n=\ell=\abar/k$ between consecutive impurities is the same for all $n$,
and
the model describes a regular lattice of point scatterers.

The continuum limit corresponds to a high density of weak scatterers.
In this limit, (\ref{eq:DefW1}) and (\ref{eq:DefW2}) approximate
two correlated Gaussian white noises such that
\begin{eqnarray}
&&\mean\left( W_1(x) \right) = \wbar/\ell = k \wbar / \abar, \nonumber\\
&&\mean\left( W_2(x) \right) = \overline{v}/\ell = k^2\ubar / \abar,
\nonumber\\
&& \text{Cov} \left( W_i(x)\, W_j(x') \right)
= \frac{1}{\ell} \, C_{ij}\, \delta(x-x')
= \frac{k}{\abar} \, C_{ij}\, \delta(x-x'),
\end{eqnarray}
where we have introduced the notation
$\text{Cov}(x\,y)=\mean(x\,y)-\mean(x)\,\mean(y)$.
The non-negative covariance matrix is
\begin{equation}
C = \pmatrix{ \dww & D_{wv} \cr D_{vw} & D_{vv} }
= \pmatrix{ \dww & k\,\dwu \cr k\,\dwu & k^2\,\duu }.
\end{equation}
It is useful for future reference to write down explicitly~(\ref{ifp})
(or~(\ref{iwasawaEquation})) in the case $\daa=0$:
\begin{eqnarray}
\label{eq:DiffEqWhenDaaIsZero}
\left [ 2 \dww\, z^2 + 2 \dwu\, z + \frac12 \duu \right ] \deriv{f}{z} \qquad
\qquad \qquad
\nonumber \\
+\left [ \abar \,(1+z^2) + 2(\dww-\wbar)\,z -\ubar \right] f(z) = j.
\end{eqnarray}

\subsection{Monolithic scalar disorder (Sec.~\ref{sec:sca})}
\label{subsec:4.1QM}

We give the quantum-mechanical interpretation of the results derived in
the case where $\daa = \dww=0$.
It is convenient to introduce the parameter
\begin{equation}
\sigma^2:=D_{vv}/\ell=k^3\duu/\abar,
\end{equation}
in terms of which
the two parameters of~(\ref{bxdef}) read
\begin{equation}
\beta=\ell\,(\sigma^2/2)^{1/3} \quad\text{and}\quad
\mu=\ell\sqrt{(\wbar/\ell)^2+\overline{v}/\ell-k^2}.
\end{equation}
We recover the result of Halperin~\cite{Ha,LGP}:
\begin{equation}
\frac1\ell\, \Omega = \left(\frac{\sigma^2}{2}\right)^{1/3} \,
G\left( \frac{(\mathbb{E}(W_1))^2+\mathbb{E}(W_2)-k^2}{(\sigma^2/2)^{2/3}}
\right),
\end{equation}
where the scaling function $G(x)$ is given by (\ref{gsca}).

\subsection{Monolithic supersymmetric disorder (Sec.~\ref{sec:susy})}
\label{subsec:4.2QM}

For the case $\daa=0 = \duu=0$, we introduce the new parameter $g:=\dww/\ell$
in terms of which
the parameters in~(\ref{parsus}) may be expressed as
$\nu=\mathbb{E}(W_1)/g=\wbar/\dww$ and
$x=\frac1g\sqrt{\overline{v}/\ell-k^2}$. Then
\begin{equation}
\frac1\ell\, \Omega = g \,
G\left( \frac{\sqrt{\mathbb{E}(W_2)-k^2}}{g} \right),
\end{equation}
where $G(x)$ is now given by (\ref{gsus}).
We have thus recovered the result of~\cite{BCGL}.

\subsection{Monolithic distance disorder (Sec.~\ref{sec:dis})}
\label{subsec:4.3QM}

In this case, the random variable $\a$ is not necessarily positive. We may
nevertheless give a quantum-mechanical interpretation in terms of random
point-like scatterers by
expressing the phases $\alpha_n$ as a sum of two parts:
a deterministic positive part $\abar$ describing the evolution of the wave
function in
the interval between two consecutive impurities separated by a
distance $\ell$, and a random part $\da_n$ modeling the effect of some
point-like scatterer.
The physical interpretation of the latter is as follows:
an electron incident on such a scatterer only experiences forward
scattering and receives the phase $\da_n$.
With this interpretation, only the averaged
phase carries an energy dependence: $\abar=k\ell$ while $\daa$ is
independent of $k$.
The two parameters (\ref{pardis}) entering the scaling function~(\ref{gdis})
take the form
\begin{equation}
\lambda=\frac{\overline{v}-2k^2\ell}{2k\daa}
\quad\text{and}\quad
x=\frac1{2k\daa}\sqrt{(2k\wbar)^2+\overline{v}^2}.
\end{equation}

\subsection{Mixed case (Sec.~\ref{sec:pot})}

The study of Equation~(\ref{eq:HamiltonianMixed}) when the two Gaussian white
noises $W_1$ and $W_2$ are
uncorrelated ($\dwu=0$) has been carried out by Hagendorf and Texier
in~\cite{HagTex08}.
The correlated case ($\dwu\neq0$) was also considered
in Appendix~A of that paper.
Following the idea proposed there,
we introduce a linear combination of the two noises:
\begin{eqnarray}
&& \phi(x) := - W_1(x) - \frac{ \dwu }{2\, \dww},\nonumber\\
&& V(x) := \left( W_2(x) - \frac{\ubar}{\abar}\right)
- \frac{ \dwu }{ \dww}
\left( W_1(x) - \frac{\wbar}{\abar} \right).
\end{eqnarray}
From now on we set $E=k^2=1$ for simplicity.
We define the three parameters~$\nu$, $g$ and $\sigma^2$ by
\begin{eqnarray}
\label{eq:DefMu}
&& \nu g = -\frac{ \wbar }{ \abar } - \frac{ \dwu }{2\dww},\\
\label{eq:DefG}
&& g = \frac{\dww}{\abar}, \\
\label{eq:DefSigma}
&& \sigma^2 = \frac{1}{\abar}\, \frac{ \duu\dww-\dwu^2 }{\dww}.
\end{eqnarray}
We may check that the two Gaussian noises $V$ and $\phi$ are
{\it uncorrelated}:
\begin{equation}
\mean(\phi(x)\,V(x'))=0,
\end{equation}
and characterized by
\begin{eqnarray}
&&\mean\left( \phi(x) \right) =\nu g, \nonumber\\
&&\mean\left( V(x) \right) =0,\nonumber\\
&&\text{Cov}\left( \phi(x)\,\phi(x') \right) = g\, \delta(x-x'),\nonumber\\
&&\mean\left( V(x)\,V(x') \right) = \sigma^2\, \delta(x-x').
\end{eqnarray}
By using the identity
\begin{equation}
W_1(x)^2-W_1'(x)+W_2(x)=\phi(x)^2+\phi'(x)+V(x)+E_0,
\end{equation}
with
\begin{equation}
E_0 = \frac{\ubar}{\abar}
- \frac{ \wbar \, \dwu}{ \abar\, \dww }
- \left(\frac{ \dwu }{2\dww }\right)^2,
\end{equation}
we may rewrite the Schr\"odinger equation
(\ref{eq:HamiltonianMixed})
in terms of the new random functions
\begin{equation}
\label{eq:HamiltonianMixed2}
H = -\deriv{^2}{x^2} + \phi(x)^2 + \phi'(x) + V(x) + E_0
=: \widetilde{H} + E_0.
\end{equation}
The spectral and localization properties of the
Hamiltonian $\widetilde{H}$ were analysed in~\cite{HagTex08}. The analysis
uses the Riccati variable
\begin{equation}
\label{eq:D21}
\tilde{z}(x) := \frac{\psi'(x)}{\psi(x)} - \phi(x).
\end{equation}
From the Schr\"odinger equation $\widetilde{H}\psi=\widetilde{E}\psi$,
where $\widetilde{E}:=E-E_0$, we see
that $\tilde{z}(x)$ obeys the Stratonovich stochastic differential equation
\begin{eqnarray}
\deriv{}{x} \tilde{z}(x) &=& - \widetilde{E}- \tilde{z}(x)^2 - 2
\tilde{z}(x)\,\phi(x) + V(x)\nonumber\\
&\eqlaw&
- \widetilde{E} -2g\nu\, \tilde{z}(x) - \tilde{z}(x)^2
+\sqrt{4g\,\tilde{z}(x)^2+\sigma^2}\,\eta(x),
\end{eqnarray}
where $\eta(x)$ is a normalized Gaussian white noise.
This equation leads to a Fokker-Planck equation for
the distribution of the Riccati process $\tilde{z}(x)$. Its limit probability
density $\tilde{f}$
satisfies the inhomogeneous first-order differential equation
\begin{equation}
\label{eq:DistributionRiccatiHT2008}
N(\widetilde{E}) =
\left[ \widetilde{E} + 2(\nu+1)g\,\tilde{z} + \tilde{z}^2
+ \left(2g\,\tilde{z}^2+ \frac{\sigma^2}{2}\right)\deriv{}{\tilde{z}}\right]
\tilde{f}(\tilde{z}),
\end{equation}
where the current $N(\widetilde{E})$ represents the integrated density of
states
per unit length of the quantum Hamiltonian $\widetilde{H}$ defined
in~(\ref{eq:HamiltonianMixed2}).
This equation is easily solved; its solution was analysed in great
detail by Hagendorf and Texier~\cite{HagTex08}.
The relation with (\ref{eq:DiffEqWhenDaaIsZero}) can be established by
noting that the Riccati variable~$z$ introduced in
Sec.~\ref{disorderedSubsection} of the
present article can be expressed in the~form
\begin{equation}
z(x)=\frac{\psi'(x)}{\psi(x)}+W_1(x).
\end{equation}
A comparison with (\ref{eq:D21}) shows that
\begin{equation}
z = \tilde{z} - \frac{\dwu}{2\dww}.
\end{equation}
With the help of this observation we can check that
(\ref{eq:DistributionRiccatiHT2008}) coincides with~(\ref{eq:DiffEqWhenDaaIsZero})
up to a constant term $\dwu$ shifting the energy.
This additive constant originates
in the definition of a continuum limit for the double impurity model,
and in the interpretation of that limit in terms of correlated Gaussian white
noises.
More precisely, it is related to the choice of the order of the $\delta$-peaks
carrying the weights $u_n$ and $w_n$,
i.e., the sign of $\varepsilon$ in~(\ref{eq:DefW1}),~(\ref{eq:DefW2}).
Indeed, if we consider the $\varepsilon\to0^-$ limit in the latter equations,
i.e., we reverse the order of the last two matrices
of the Iwasawa decomposition~(\ref{IwasawaDecomposition}),
then~(\ref{ifp}) leads to~(\ref{eq:DiffEqWhenDaaIsZero})
up to a constant term $2\dwu$.
The result for the continuum model~\cite{HagTex08}
is therefore given by the arithmetic mean
of the discrete expressions in the $\varepsilon\to0^+$ and $\varepsilon\to0^-$
limits.

Recall that the integrated density of states per unit length measures the
density of zeros of the wave function $\psi(x)$, that is also the
current of the Riccati variable through $\mathbb{R}$
(see below~(\ref{omegadef})).
Hence
\begin{equation}
j=\abar\,N(\widetilde{E}).
\end{equation}
In~\cite{HagTex08}, the characteristic function
\begin{equation}
\Omega(\widetilde{E}+\ii0^+)=\gamma(\widetilde{E})-\ii\pi\,N(\widetilde{E})
\end{equation}
associated with the Hamiltonian
(\ref{eq:HamiltonianMixed2}) was found in
terms of a (Tricomi) confluent hypergeometric function:
\begin{equation}
\label{eq:OmegaChristian}
\Omega(\widetilde{E}+\ii0^+) = - \nu g
+ g \, \xi
\left[
1 - 2
\frac{U'\left(\frac{\nu+1}{2}+\theta,\nu+1; \xi\right)}
{U\left(\frac{\nu+1}{2}+\theta,\nu+1; \xi\right)}
\right],
\end{equation}
where
\begin{equation}
\label{eq:9}
\xi :=-\frac{\ii}{2} \sqrt{\frac{\sigma^2}{g^3}}
\quad
\mbox{and}
\quad
\theta :=
\frac\ii2
\left(
\frac14\sqrt{\frac{\sigma^2}{g^3}}-\frac{\widetilde{E}}{\sqrt{\sigma^2 g}}
\right).
\end{equation}
(Note that Equation~(102) of~\cite{HagTex08} contains a misprint, namely a
superfluous
$\nu+1$ in the denominator.)
In terms of the Whittaker function, we have
\begin{equation}
U(a,b;z)=z^{-b/2}\e^{z/2}\,W_{-a+{b}/{2},(b-1)/2}(z),
\end{equation}
and hence
\begin{equation}
\label{eq:OmegaJeanMarc}
\Omega(\widetilde{E}+\ii0^+) =
g \, \left(
1 - 2\xi\,
\frac{ W_{-\theta,\frac\nu2}'(\xi)
}{W_{-\theta,\frac\nu2}(\xi)}
\right).
\end{equation}
Equation~(\ref{eq:OmegaJeanMarc}) has precisely the same structure
as~(\ref{gwhi}).
With the help of~(\ref{eq:DefMu})--(\ref{eq:DefSigma}),
we easily identify $\xi$ with (\ref{xpartdef}), up to a complex conjugation,
and $\nu/2$ with (\ref{eq:ParameterM}).
The parameter $\theta$, where $\widetilde{E}=E-E_0=1-E_0$, coincides with
(\ref{eq:ParameterL}), up to a shift of $E$ by $\dwu$.

Finally, let us point out that, interestingly, the limit of
correlated noises for (\ref{eq:HamiltonianMixed}) with
$\dwu=\pm\sqrt{\dww\,\duu}$ maps to a
quantum Hamiltonian (\ref{eq:HamiltonianMixed2}) with $V(x)=0$; this is the
supersymmetric model
studied earlier by Bouchaud {\em et al.} in~\cite{BCGL}.
This explains why (\ref{gsus}) and (\ref{eq:ggpdfc}) coincide.
Note, however, that the limit $\sigma^2\to0$ is highly singular, as one
would expect from the expressions obtained for the characteristic
function,
(\ref{eq:OmegaChristian}) and (\ref{eq:OmegaJeanMarc}).
This limit was carefully analysed in~\cite{HagTex08}.

\section{Summary}
\label{sec:disc}

In this work we have carried out a systematic study
of the complex characteristic exponent $\Omega$
of products of arbitrary random matrices close to the identity in the group
$\text{SL}(2,{\mathbb R})$.
Our main result is that the characteristic exponent
admits a scaling form in the continuum limit
where the three mean Iwasawa parameters $\abar$, $\wbar$,~$\ubar$
and the six elements of their covariance matrix are small and comparable.
This result is fully general;
it provides a unified approach for many results
found in earlier works for specific cases.

In the above continuum scaling regime,
the complex characteristic exponent~$\Omega$
is a homogeneous function of its nine variables with degree one.
Interestingly, the corresponding scaling function is (essentially) given
by the logarithmic derivative of a special function of mathematical analysis.
The general case involves Gauss's hypergeometric function,
while particular cases involve other special functions
(Airy, Bessel, Whittaker, elliptic).
All these special functions (and many others)
can be obtained from the confluent hypergeometric one
by limiting procedures~\cite[Table 13.6, p.~509]{as}.
In the present framework, however,
the type of special function pertaining to each case
is entirely dictated by the singularities
of the integrating factor $H(y)$ at the zeros of the polynomial $Q(y)$
---the complexified diffusion coefficient $\sigma^2(z)$.
These singularities are in turn set by the multiplicities of the latter zeros,
according to the correspondence given in Table~\ref{I} at the end of the introduction.
As the present analysis of the Lyapunov exponent holds in full generality,
at least for matrices in $\text{SL}(2,{\mathbb R})$,
this table gives a classification
of the possible types of one-dimensional continuous disordered models,
viewed as continuum limits of discrete models described by products of random matrices.

The line of thought of the present work can be extended to products
of more general $2\times2$ matrices.
We have shown explicitly in Sec.~\ref{sec:zd} how the case of matrices
close to the identity in $\text{GL}(2,{\mathbb R})$,
whose determinant is not unity, can also be handled efficiently.
The same holds true for complex matrices in $\text{GL}(2,{\mathbb C})$.
Finally, the scaling forms of the characteristic exponent put forward in this work
may also hold for random matrices which are weakly non-commuting,
albeit not close to the identity.
One explicit example is the tight-binding Anderson model near its band edges,
whose continuum limit coincides with the Halperin model,
i.e., our monolithic scalar disorder.
Extending the present analysis to larger matrices however remains a formidable challenge.

\section*{Acknowledgments}

This work was supported by {\em Triangle de la Physique}.
We are pleased to thank Ph. Bougerol for illuminating discussions.

\appendix
\section{Derivation of Equation~(\ref{transformedIfp})}
\label{transformAppendix}

In order to derive Equation~(\ref{transformedIfp}),
it will be convenient to work with the following equivalent form of the
integrated Fokker-Planck equation (\ref{ifp}):
\begin{equation}
\frac{1}{2} \,{\mathbf g}(z) \cdot \frac{\d}{\d z} \left [ f(z)
\,\boldsymbol{\sigma^2} {\mathbf g}(z) \right ] + f(z) \left \{
\AvParam \cdot {\mathbf g(z)} + \frac{1}{2} \,{\mathbf c} \cdot
\left [ {\mathbf g}(z) \times {\mathbf g}'(z) \right ] \right \} = j.
\label{iwasawaEquation}
\end{equation}

Let $R$ be a large positive number, destined to tend to infinity. Multiply the
above
equation by $1/(y-z)$
and integrate over $z$ from $-R$ to $R$:
\begin{eqnarray}
&&\frac{1}{2} \,\int_{-R}^R \frac{\d z}{y-z} {\mathbf g}(z) \cdot \frac{\d}{\d
z} \left [ f(z) \,\boldsymbol{\sigma^2} {\mathbf g}(z) \right ]
\nonumber\\
&&+\int_{-R}^R \frac{\d z}{y-z} f(z) \left \{
\AvParam \cdot {\mathbf g(z)} + \frac{1}{2} \,{\mathbf c} \cdot
\left [ {\mathbf g}(z) \times {\mathbf g}'(z) \right ] \right \}
= \int_{-R}^R \frac{j \,\d z}{y-z}.
\label{intermediateIwasawa}
\end{eqnarray}
There follows a discussion of each of the terms appearing in this expression.

Let us deal firstly with the term on the right-hand side. Consider the closed
semicircular contour $C$ of radius $R$
centered on the origin in the lower half of the complex $z$-plane. By the
Residue Theorem,
\begin{equation}
\int_C \frac{\d z}{z-y} = 2 \pi \ii,
\end{equation}
and so
\begin{equation}
\int_{R}^{-R} \frac{\d z}{z-y} + \int_{-\pi}^0 \frac{\ii \,R \e^{\ii \theta} \d
\theta}{R \e^{\ii \theta}-\ii \,\text{Im} \,y} = 2 \pi\ii.
\end{equation}
We deduce
\begin{equation}
\int_{-R}^R \frac{j \,\d z}{y-z}\comport{\longrightarrow}{R\to\infty}\ii \pi j.
\label{log}
\end{equation}

Next, consider the left-hand side. Since the
entries of ${\mathbf g}(z)$ are
polynomials of degree at most two, Taylor's Theorem says that
\begin{equation}
{\mathbf g} (z) = {\mathbf g} (y) - (y-z) \,{\mathbf g}'(y) +
\frac{(y-z)^2}{2} \,{\mathbf g}''(y).
\label{taylorIdentity}
\end{equation}
The first term on the left-hand side of (\ref{intermediateIwasawa})
may therefore be developed to yield
\begin{eqnarray}
&&\frac{1}{2} \int_{-R}^R \frac{\d z}{y-z} {\mathbf g}(z) \cdot \frac{\d}{\d
z} \left [ f(z) \,\boldsymbol{\sigma^2} {\mathbf g}(z) \right ]
\nonumber\\
&=& {\mathbf g}(y) \cdot \frac{1}{2} \,\int_{-R}^R \frac{\d z}{y-z}
\frac{\d}{\d z} \left [ f(z) \,\boldsymbol{\sigma^2} {\mathbf g}(z) \right ]
- {\mathbf g}'(y) \cdot \frac{1}{2} \,\int_{-R}^R \d z \frac{\d}{\d z} \left
[ f(z) \,\boldsymbol{\sigma^2} {\mathbf g}(z) \right ]
\nonumber\\
&+& \frac{{\mathbf g}''(y)}{2} \cdot \frac{1}{2} \,\int_{-R}^R \d z (y-z)
\frac{\d}{\d z} \left [ f(z) \,\boldsymbol{\sigma^2} {\mathbf g}(z) \right ]
\label{intermediateFirstRHSTerm}
\end{eqnarray}
The right-hand side of this expression is the sum of three terms; we deal with
each of them in turn. For the first term, integration by parts gives
\begin{equation}
{\mathbf g}(y) \cdot \frac{1}{2}\, \left \{ \frac{1}{y-z} f(z)
\,\boldsymbol{\sigma^2} {\mathbf g}(z) \Bigl |_{-R}^R -
\int_{-R}^R \frac{\d z}{(y-z)^2} f(z) \,\boldsymbol{\sigma^2} {\mathbf g}(z)
\right \}.
\end{equation}
We now let $R \to \infty$ to obtain
\begin{eqnarray}
&-&{\mathbf g}(y) \cdot \frac{1}{2}\,\int_{-\infty}^\infty \frac{\d z}{(y-z)^2}
f(z) \,\boldsymbol{\sigma^2} {\mathbf g}(z)
\nonumber\\
=&-&\frac{1}{2}\, {\mathbf g}(y) \cdot \boldsymbol{\sigma^2} {\mathbf g}(y)
\int_{-\infty}^\infty \frac{\d z}{(y-z)^2} f(z)
+ \frac{1}{2}\, {\mathbf g}(y) \cdot \boldsymbol{\sigma^2} {\mathbf g}'(y)
\int_{-\infty}^\infty \frac{\d z}{y-z} f(z)
\nonumber\\
&-&\frac{1}{4}\, {\mathbf g}(y) \cdot \boldsymbol{\sigma^2} {\mathbf g}''(y)
\int_{-\infty}^\infty \d z f(z).
\end{eqnarray}
Recalling the definition of the transform $F$, we conclude that the first
term on the right-hand side of (\ref{intermediateFirstRHSTerm})
yields
\begin{eqnarray}
&&{\mathbf g}(y) \cdot \frac{1}{2} \,\int_{-R}^R \frac{\d z}{y-z} \frac{\d}{\d
z} \left [ f(z) \,\boldsymbol{\sigma^2} {\mathbf g}(z) \right ]
\nonumber\\
&&\comport{\longrightarrow}{R\to \infty}
\frac{1}{2} \,{\mathbf g}(y) \cdot \frac{\d}{\d y} \left [ F(y)
\,\boldsymbol{\sigma}^2 {\mathbf g}(y) \right ]
-\frac{1}{4}\, {\mathbf g}(y) \cdot \boldsymbol{\sigma^2} {\mathbf g}''(y).
\end{eqnarray}

The second term on the right-hand side of
(\ref{intermediateFirstRHSTerm}) can be integrated immediately:
\begin{equation}
- {\mathbf g}'(y) \cdot \frac{1}{2} \,\int_{-R}^R \d z \frac{\d}{\d z} \left
[ f(z) \,\boldsymbol{\sigma^2} {\mathbf g}(z) \right ]
= - {\mathbf g}'(y) \cdot \frac{1}{2} \,f(z) \,\boldsymbol{\sigma^2} {\mathbf
g}(z) \Bigl |_{-R}^R
\comport{\longrightarrow}{R \to \infty} 0
\end{equation}
by virtue of the Rice formula (\ref{riceFormula}).

The third term on the right-hand side of
(\ref{intermediateFirstRHSTerm}) is the most delicate:
\begin{eqnarray}
&&\frac{{\mathbf g}''(y)}{2} \cdot \frac{1}{2} \,\int_{-R}^R \d z (y-z)
\frac{\d}{\d z} \left [ f(z) \,\boldsymbol{\sigma^2} {\mathbf g}(z) \right ]
\nonumber\\
&=&\frac{{\mathbf g}''(y)}{4} \cdot \boldsymbol{\sigma^2} {\mathbf
g}(y)\,\int_{-R}^R \d z (y-z) f'(z)
\nonumber\\
&-&\frac{{\mathbf g}''(y)}{4} \cdot \boldsymbol{\sigma^2} {\mathbf g}'(y)
\,\int_{-R}^R \d z (y-z) \frac{\d}{\d z} \left [ f(z) \,(y-z) \right ]
\nonumber\\
&+&\frac{{\mathbf g}''(y)}{8} \cdot \boldsymbol{\sigma^2} {\mathbf g}''(y)
\,\int_{-R}^R \d z (y-z) \frac{\d}{\d z} \left [ f(z) \,(y-z)^2 \right ].
\label{thirdTermIntermediate}
\end{eqnarray}
We may use integration by parts for the first and second terms on the
right-hand side of this expression. We find
\begin{equation}
\frac{{\mathbf g}''(y)}{4} \cdot \boldsymbol{\sigma^2} {\mathbf
g}(y)\,\int_{-R}^R \d z (y-z) f'(z)
\comport{\longrightarrow}{R \to \infty}
\frac{{\mathbf g}''(y)}{4} \cdot \boldsymbol{\sigma^2} {\mathbf g}(y)
\end{equation}
and, by using (\ref{rice1}),
\begin{eqnarray}
&&- \frac{{\mathbf g}''(y)}{4} \cdot \boldsymbol{\sigma^2} {\mathbf g}'(y)
\,\int_{-R}^R \d z (y-z) \frac{\d}{\d z} \left [ f(z) \,(y-z) \right ]
\nonumber\\
&&\comport{\longrightarrow}{R \to \infty}
- \frac{{\mathbf g}''(y)}{4} \cdot \boldsymbol{\sigma^2} {\mathbf g}'(y)
\,\int_{-\infty}^\infty \d z (y-z) f(z)
\end{eqnarray}

Finally, for the third term on the right-hand side of
(\ref{thirdTermIntermediate}):
\begin{eqnarray}
&&\frac{{\mathbf g}''(y)}{8} \cdot \boldsymbol{\sigma^2} {\mathbf g}''(y)
\,\int_{-R}^R \d z (y-z) \frac{\d}{\d z} \left [ f(z) \,(y-z)^2 \right ]
\nonumber\\
&=&
\frac{{\mathbf g}''(y)}{8} \cdot \boldsymbol{\sigma^2} {\mathbf g}''(y) \,\left
\{ y^2 \int_{-R}^R \d z (y-z) f'(z) -2 y \int_{-R}^R \d z (y-z) \frac{\d}{\d
z} \left [ z f(z) \right ] \right.
\nonumber\\
&&{\hskip 150pt}\left. + \int_{-R}^R \d z (y-z) \frac{\d}{\d z} \left [ z^2
f(z) \right ]
\right \}
\end{eqnarray}
By using integration by parts and~(\ref{rice1})--(\ref{rice2}), we
obtain
\begin{eqnarray}
&&\frac{{\mathbf g}''(y)}{8} \cdot \boldsymbol{\sigma^2} {\mathbf g}''(y)
\,\int_{-R}^R \d z (y-z) \frac{\d}{\d z} \left [ f(z) \,(y-z)^2 \right ]
\nonumber\\
&&\comport{\longrightarrow}{R \to \infty}
\frac{{\mathbf g}''(y)}{8} \cdot \boldsymbol{\sigma^2} {\mathbf g}''(y) \,\left
\{ y^2 -2 y \int_{-\infty}^\infty z f(z) \,\d z
- \int_{-\infty}^\infty z \frac{\d}{\d z} \left [ z^2 f(z) \right ] \,\d z
\right \}.
\end{eqnarray}
This completes our discussion of the first term on the left-hand side of
(\ref{intermediateIwasawa}).

The second term on that left-hand side is easier to deal with since no
derivative of $f$ appears and hence there is no need to
integrate by parts. By using (\ref{taylorIdentity}), this term is
easily developed to yield
\begin{eqnarray}
&&\int_{-R}^R \frac{\d z}{y-z} f(z) \left \{
\AvParam \cdot {\mathbf g}(z) + \frac{1}{2} \,{\mathbf c} \cdot
\left [ {\mathbf g}(z) \times {\mathbf g}'(z) \right ] \right \}
\nonumber\\
&&\comport{\longrightarrow}{R \to \infty}
\left \{
\AvParam \cdot {\mathbf g} (y) + \frac{1}{2} \,{\mathbf c} \cdot
\left [ {\mathbf g}(y) \times {\mathbf g}'(y) \right ] \right \} F(y) -
\AvParam \cdot {\mathbf g}'(y) - \frac{1}{2} {\mathbf c} \cdot
\left [ {\mathbf g} (y) \times {\mathbf g}''(y)
\right ]
\nonumber\\
&&\qquad+ \left \{
\frac{1}{2} \AvParam \cdot {\mathbf g''(y)} + \frac{1}{4}
\,{\mathbf c} \cdot \left [ {\mathbf g}'(y) \times {\mathbf g}''(y) \right ]
\right \} \int_{-\infty}^{\infty} (y-z) f(z)\,\d z.
\end{eqnarray}

Equation~(\ref{transformedIfp}) follows after putting all these
results together and making use of (\ref{simplifiedLyapunov}) for the
Lyapunov exponent.


\begin{thebibliography}{99}

\bibitem{BL} Bougerol, P., Lacroix, J.: {\it Products of Random Matrices with Applications to Schr\"{o}\-dinger Operators.} Birha\"{u}ser, Basel (1985).

\bibitem{cpv} Crisanti, A., Paladin, G., Vulpiani, A.: {\it Products of Random Matrices in Statistical Physics.} Springer, Berlin (1992).

\bibitem{Lu} Luck, J.M.: {\it Syst\`emes d\'esordonn\'es unidimensionnels.} Collection Al\'ea-Saclay (1992).

\bibitem{pendry} Pendry, J.B.: Symmetry and transport of waves in one-dimensional disordered systems. Adv. Phys. {\bf 43}, 461-542 (1994).

\bibitem{Dy} Dyson, F.J.: The dynamics of a disordered linear chain. Phys. Rev. {\bf 92}, 1331-1338 (1953).

\bibitem{Sch} Schmidt, H.: Disordered One-Dimensional Crystals. Phys. Rev. {\bf 105}, 425-441 (1957).

\bibitem{Bor63} Borland, R.E.: The nature of the electronic states in disordered one-dimensional systems. Proc. Roy. Soc. London {\bf A 274}, 529-545 (1963).

\bibitem{horvai} Horvai, P.: Lyapunov exponent for inertial particles in the 2D Kraichnan model as a problem of Anderson localization with complex valued potential, arXiv preprint nlin/0511023 (2005).

\bibitem{gawedzki} Gawedzki, K., Herzog, D.P., Wehr, J.: Ergodic properties of a model for turbulent dispersion of inertial particles. Commun. Math. Phys. {\bf 308}, 49-80 (2011).

\bibitem{lloyd} Lloyd, P.: Exactly solvable model of electronic states in a three-dimensional disordered Hamiltonian: non-existence of localized states. J. Phys. C {\bf 2}, 1717-1725 (1969).

\bibitem{theoh} Nieuwenhuizen, Th.M.: Exact solutions for spectra and Green's functions in random one-dimensional systems. Physica A {\bf 125}, 197-236 (1984).

\bibitem{theo} Nieuwenhuizen, Th.M.: Exact electronic spectra and inverse localization lengths in one-dimensional random systems: I. Random alloy, liquid metal and liquid alloy. Physica A {\bf 120}, 468-514 (1983).

\bibitem{theopl} Nieuwenhuizen, Th.M.: Exactly soluble diluted random one-dimensional lattices. Phys. Lett. A {\bf 103}, 333-336 (1984).

\bibitem{BaLu} Barnes, C., Luck, J.M.: The distribution of the reflexion phase of disordered conductors. J. Phys. A: Math. Theor. {\bf 23}, 1717-1734 (1990).

\bibitem{NL} Nieuwenhuizen, Th.M., Luck, J.M.: Exactly soluble random field Ising models in one dimension. J. Phys. A: Math. Theor. {\bf 19}, 1207-1227 (1986).

\bibitem{LN} Luck, J.M., Nieuwenhuizen, Th.M.: Correlation function of random-field Ising chains: is it Lorentzian or not? J. Phys. A: Math. Theor. {\bf 22}, 2151-2180 (1989).

\bibitem{FNT} Funke, M., Nieuwenhuizen, Th.M., Trimper, S.: Exact solutions for Ising chains in a random field. J. Phys. A: Math. Theor. {\bf 22}, 5097-5107 (1989).

\bibitem{LFN} Luck, J.M., Funke, M., Nieuwenhuizen, Th.M.: Low-temperature thermodynamics of random-field Ising chains: exact results. J. Phys. A: Math. Theor. {\bf 24}, 4155-4196 (1991).

\bibitem{jmspin} Luck, J.M.: Critical behavior of the aperiodic quantum Ising chain in a transverse magnetic field. J. Stat. Phys. {\bf 72}, 417-458 (1993).

\bibitem{PF} Pastur, L., Figotin, A.: {\it Spectra of Random and Almost-Periodic Operators.} Springer, Berlin (1992).

\bibitem{CaLa} Carmona, R., Lacroix, J.: {\it Spectral Theory of Random Schr\"{o}dinger Operators.} Birk\-h\"auser, Boston (1990).

\bibitem{MTW} Marklof, J., Tourigny, Y., Wo{\l}owski, L.: Explicit invariant measures for products of random matrices. Trans. Amer. Math. Soc. {\bf 360}, 3391-3427 (2008).

\bibitem{CTT1} Comtet, A., Texier, C., Tourigny, Y.: Products of Random Matrices and Generalised Quantum Point Scatterers. J. Stat. Phys. {\bf 140}, 427-466 (2010).

\bibitem{CTT2} Comtet, A., Texier, C., Tourigny, Y.: Supersymmetric quantum mechanics with L\'evy disorder in one dimension. J. Stat. Phys. {\bf 145}, 1291-1323 (2011).

\bibitem{CTT3} Comtet, A., Texier, C., Tourigny, Y.: Lyapunov exponents, one-dimensional Anderson localisation and products of random matrices. Preprint arXiv:1207.0725. To appear in a special issue of J. Phys. A: Math. Theor. (March 2013).

\bibitem{dmp} Derrida, B., Mecheri, K., Pichard, J.L.: Lyapounov exponents of products of random matrices: weak disorder expansion. Application to localisation. J. Phys. (France) {\bf 48}, 733-740 (1987).

\bibitem{SSB} Sadel, C., Schulz-Baldes, H.: Random Lie group actions on compact manifolds: a perturbative analysis. Ann. Prob. {\bf 36}, 2224-2257 (2010).

\bibitem{zd} Zanon, N., Derrida, B.: Weak Disorder Expansion of Liapunov Exponents in a Degenerate Case. J. Stat. Phys. {\bf 50}, 509-528 (1988).

\bibitem{ChLe} Chamayou, J.F., Letac, G.: Explicit stationary distributions for compositions of random functions and products of random matrices. J. Theor. Prob. {\bf 4}, 3-36 (1991).

\bibitem{Fu} Furstenberg, H.: Noncommuting random products. Trans. Amer. Math. Soc. {\bf 108}, 377-428 (1963).

\bibitem{ADK} Albeverio, S., Dabrowski, L., Kurasov, P.: Symmetries of Schr\"{o}dinger operators with point interactions. Lett. Math. Phys. {\bf 45}, 33-47 (1998).

\bibitem{Se} \u{S}eba, P.: The generalized point interaction in one dimension. Czech. J. Phys. {\bf 36}, 667-673 (1986).

\bibitem{KP} Kronig, R. de L., Penney, W.G.: Quantum mechanics of electrons in crystal lattices. Proc. Roy. Soc. London {\bf A 130}, 499-513 (1931).

\bibitem{LGP} Lifshits, I.M., Gredeskul, S.A., Pastur, L.A.: {\it Introduction to the Theory of Disordered Systems.} Wiley, New York (1988).

\bibitem{FL} Frisch, H.L., Lloyd, S.P.: Electron Levels in a One-Dimensional Random Lattice. Phys. Rev. {\bf 120}, 1175-1189 (1960).

\bibitem{Ha} Halperin, B.I.: Green's Functions for a Particle in a One-Dimensional Random Potential. Phys. Rev. {\bf 139}, A104-A117 (1965).

\bibitem{HJ} Herbert, D.C., Jones, R.: Localized states in disordered systems. J. Phys. C {\bf 4}, 1145-1161 (1971).

\bibitem{Th} Thouless, D.J.: A relation between the density of states and range of localization for one-dimensional random systems. J. Phys. C: Solid State Phys. {\bf 5}, 77-81 (1972).

\bibitem{iwasawa} Iwasawa, K.: On some types of topological groups. Annals of Math. {\bf 50}, 507-558 (1949).

\bibitem{NS} Nechaev, S.K., Sinai, Ya.G.: Limiting-type theorem for conditional distributions of products of independent unimodular $2 \times 2$ matrices. Bol. Soc. Brasil. Mat. (N.S.) {\bf 21}, 121-132 (1991).

\bibitem{nist} Olver, F.W.J., Lozier, D.W., Boisvert, R.F., Clark, C.W. (eds.): {\it NIST Handbook of Mathematical Functions.} Cambridge University Press, Cambridge (2010).

\bibitem{BCGL} Bouchaud, J.P., Comtet, A., Georges, A., Le Doussal, P.: Classical diffusion of a particle in a one-dimensional random force field. Ann. Phys. {\bf 201}, 285-341 (1990).

\bibitem{HagTex08} Hagendorf, C., Texier, C.: Breaking supersymmetry in a one-dimensional random Hamiltonian. J. Phys. A: Math. Theor. {\bf 41}, 405302 (2008).

\bibitem{rice} Rice, S.O.: Mathematical Analysis of Random Noise. Bell System Tech. J. {\bf 23}, 282-332 (1944).

\bibitem{as} Abramowitz, M., Stegun, I.A.: {\it Handbook of Mathematical Functions.} Dover, New York (1974).

\bibitem{sulem} Sulem, P.L.: Total reflection of a plane wave from a semi-infinite, one-dimensional random medium: Distribution of the phase. Physica {\bf 70}, 190-208 (1973).

\bibitem{dgedge} Derrida, B., Gardner, E.J.: Lyapounov exponent of the one dimensional Anderson model: weak disorder expansions. J. Phys. (France) {\bf 45}, 1283-1295 (1984).

\bibitem{tit} Erd\'elyi, A. (ed.): {\it Tables of Integral Transforms.} McGraw-Hill, New-York (1954).

\bibitem{htf} Erd\'elyi, A. (ed.): {\it Higher Transcendental Functions.} McGraw-Hill, New-York (1953).

\bibitem{bg} Bouchaud, J.P., Georges, A.: Anomalous diffusion in disordered media: Statistical mechanisms, models and physical applications. Phys. Rep. {\bf 195}, 127-293 (1990).

\bibitem{gr} Gradshteyn, I.S., Ryzhik, I.M.: {\it Table of Integrals, Series, and Products.} Academic, New York (1965).

\bibitem{FJ} Franchi, J., Le Jan, Y.: {\it Hyperbolic Dynamics and Brownian Motions - An Introduction.} Oxford Mathematical Monographs. Oxford University Press, Oxford (2012).

\bibitem{CM} Comtet, A., Monthus, C.: Diffusion in a one-dimensional random medium and hyperbolic Brownian motion. J. Phys. A: Math. Theor. {\bf 29}, 1331-1345 (1996).

\bibitem{MY} Yor, M.: On some exponential functionals of Brownian motion. Adv. Appl. Probab. {\bf 28}, 509-531 (1992).

\bibitem{AndThoAbrFis80} Anderson, P.W., Thouless, D.J., Abrahams, E., Fisher, D.S.: New method for a scaling theory of localization. Phys. Rev. B {\bf 22}, 3519-3526 (1980).

\bibitem{CohRotSha88} Cohen, A., Roth, Y., Shapiro, B.: Universal distributions and scaling in disordered systems. Phys. Rev. B {\bf 38}, 12125-12132 (1988).

\bibitem{deych} Deych, L.I., Lisyansky, A.A., Altschuler, B.L.: Single Parameter Scaling in One-Dimensional Localization Revisited. Phys. Rev. Lett. {\bf 84}, 2678-2681 (2000).

\bibitem{SchTit03} Schomerus, H., Titov, M.: Band-center anomaly of the conductance distribution in one-dimensional Anderson localization. Phys. Rev. B {\bf 67}, 100201(R) (2003).

\bibitem{schrader} Schrader, R., Schulz-Baldes, H., Sedrakyan, A.: Perturbative Test of Single Parameter Scaling for 1D Random Media. Annales Henri Poincar\'e {\bf 5}, 1159-1180 (2004).

\end{thebibliography}
\end{document}